\documentclass{aastex701}

\begin{document}

\title{Galaxy Spiral Arm Count vs.\ Concentration and Mass: A First Look with Euclid}

\author[0000-0002-8521-5240]{Beverly J. Smith}
\affiliation{Department of Physics and Astronomy, East Tennessee
State University, Johnson City TN 37614}
\email[show]{smithbj@etsu.edu}  

\author[]{Sydnie Bergner} 
\affiliation{Department of Physics and Astronomy, East Tennessee
State University, Johnson City TN 37614}
\email{bergner@etsu.edu}

\author[0000-0003-3856-7216]{Mark L. Giroux}
\affiliation{Department of Physics and Astronomy, East Tennessee
State University, Johnson City TN 37614}
\email{girouxm@etsu.edu}

\author[0000-0002-6490-2156]{Curtis Struck}
\affiliation{Iowa State University, Department of Astronomy, Ames IA}
\email{curt@iastate.edu}

\begin{abstract}

Using catalogued information 
from Euclid Quick Data Release 1,
we compare 2-armed and 3-armed spiral galaxies
as classified by the Euclid Zoobot software.
Two-armed galaxies have larger concentrations, lower stellar masses (M*),
and lower star formation rates (SFRs)
on average than 3-armed galaxies.
For a given M*, 2-armed galaxies have larger concentrations than 3-armed galaxies.
These trends have been seen before in nearby galaxies; with Euclid we extend
the patterns to redshifts z = 0.4 $-$ 1.
Two-armed galaxies have lower SFRs because they have lower masses;
at fixed M*, 2-armed and 3-armed galaxies have similar SFRs.
We see a bend in the concentration-log M* relation for 2-armed
galaxies at M* $\approx$ 10$^{10.3}$ M$_{\sun}$.  Above this mass, 
2-armed galaxies show 
significantly larger concentrations than their lower mass counterparts.
The observed concentrations of 2-armed galaxies decrease
with increasing redshift, perhaps from morphological K-corrections and resolution
differences.
About 60\% $-$ 70\% of Euclid
spirals 
are
2-armed, and about 15\% $-$ 20\% are 3-armed.
One-armed galaxies are rare, 
with low masses compared to 2-armed spirals.
We compare these statistics with 
Galaxy Zoo Sloan Digital Sky
Survey arm counts at low z, and tentatively with JWST 
at higher z.  We discuss these results in terms
of theoretical models of spiral arm generation and evolution,
and compare with statistics 
of grand design, multi-armed, and flocculent galaxies.
There is a need for more quantitative
measurements of arm structure
beyond arm counts provided by the Zoobot/Galaxy Zoo or the 
three standard arm classes.

\end{abstract}



\section{Introduction} 

Spiral galaxies 
can be divided into classes based on
the number of arms and the structure of the arms. Grand design
spiral galaxies have two long continuous arms, multi-armed galaxies
have three or more relatively long arms, and flocculent
galaxies have many short arm fragments
\citep{2011ApJ...737...32E, 2015ApJS..217...32B}.
Multi-armed spirals sometimes have two arms in their interior, which
branch into four or more arms further out in the disk
\citep{1982MNRAS.201.1021E, 
2004A&A...423..849G,
2015ApJS..217...32B,
2024AJ....168...12S,
2024AJ....168..264W}.
Our own Milky Way is multi-armed, with 
four outer arms 
\citep{2016AJ....151...55V,
2019ApJ...885..131R}
possibly branching from two arms in the interior 
\citep{2021A&A...654A.138M,
2023ApJ...947...54X,
2025AJ....170..243K}.
The number of arms in a galaxy and their lengths may reflect both the underlying mass
distribution of the galaxy as well as its dynamical history.

Observationally, 
the number of arms correlate with other properties of the galaxies.
Grand design galaxies tend to have earlier Hubble types
and larger concentrations than multi-armed galaxies
with the same stellar mass 
\citep{1982MNRAS.201.1021E,
2017MNRAS.471.1070B,
2020ApJ...900..150Y,
2024AJ....168...12S}.
Grand design and multi-armed galaxies populate different regions in
a concentration-stellar mass diagram,
with grand design galaxies tending to 
have larger concentrations but 
smaller stellar masses on average 
\citep{2024AJ....168...12S}.
Flocculent galaxies tend to have lower masses 
\citep{2017MNRAS.471.1070B},
lower luminosities 
\citep{1982MNRAS.201.1021E,
2013JKAS...46..141A},
smaller sizes 
\citep{1987ApJ...314....3E},
and later Hubble types
\citep{2011ApJ...737...32E, 
2013JKAS...46..141A}
than other spirals.
These observational trends provide clues to the processes that generate
and maintain spiral arms in galaxies.

Multiple mechanisms may contribute to the spiral patterns observed
in galaxies.
Gravitational instabilities 
in isolated differentially-rotating disks can produce
transient but recurring
spiral patterns 
\citep{1965MNRAS.130..125G},
which can be
enhanced by galactic shear via swing amplification 
\citep{1966ApJ...146..810J, 2012ApJ...751...44S,
2013ApJ...763...46B,
2013ApJ...766...34D, 2014PASA...31...35D, 2018MNRAS.481..185M}.
Tidal interactions between galaxies can also induce
spiral patterns in disks
\citep{1972ApJ...178..623T, 1992AJ....103.1089B, 
2008ApJ...683...94O,
2010MNRAS.403..625D, 
2011MNRAS.414.2498S,
2015ApJ...807...73O}.
Galactic bars may 
drive spiral patterns 
\citep{1976ApJ...209...53S, 1979ApJ...233..539K,
1980A&A....88..184A, 2010ApJ...715L..56S}.
Spiral density waves 
\citep{1964ApJ...140..646L, 1966PNAS...55..229L,
2016ARA&A..54..667S}
may be present in some galaxies,
and may be
long-lived under some circumstances 
\citep{1985IAUS..106..513L, 1989ApJ...338...78B,
1989ApJ...338..104B, 2016ApJ...826L..21S}.

According to classical spiral density wave theory, 
a dense concentrated bulge may stabilize
a two-armed spiral density wave pattern 
and extend its lifetime
\citep{1985IAUS..106..513L, 1989ApJ...338...78B,
1989ApJ...338..104B}
an idea supported by numerical simulations
\citep{2016ApJ...826L..21S}.
This is compatible with 
the tendency of
grand design galaxies
to have large bulges.
The morphologies of flocculent galaxies can be reproduced by
simulations of stochastic gravitational collapse in 
differentially-rotating disks 
\citep{2013ApJ...766...34D}.
These kinds of models can also produce
spiral patterns like those seen in multi-armed galaxies,
if the simulation has a large disk mass compared to the dark matter
surface density
\citep{2013ApJ...766...34D, 2018MNRAS.477.1451F}.
In these models, the arms tend to be long and well-defined.
In contrast, 
simulations with proportionally smaller disk/halo mass surface
densities tend to produce
flocculent spirals 
\citep{2013ApJ...766...34D, 2018MNRAS.477.1451F}.
Models with high
star formation rates (SFRs) and high stellar feedback
are more likely to produce flocculent morphologies
\citep{2018MNRAS.478.3793D}.
Simulations of stochastic production of spiral
arms in isolated galaxies
can also produce transient 2-armed patterns if they have 
large disk/halo masses, large bulges, and high shear rates 
\citep{2013A&A...553A..77G, 2015ApJ...808L...8D, 2018MNRAS.481..185M, 2025Galax..13..132P}.
The spiral patterns created by bar-driving 
models tend to be 2-armed 
\citep{2012MNRAS.426L..46A},
or 2-armed in the interior, branching to 4-armed in the outskirts 
\citep{2004A&A...423..849G}.
Observationally,
barred galaxies tend to have two arms in the inner region, branching
to form multiple arms further out 
\citep{2024AJ....168..264W}.
Branching may be associated with orbital resonances
\citep{1997A&A...323..762P, 1995ApJ...445..591E,
2003ApJ...596..220C}.
Barred galaxies tend to have stronger spiral arms than unbarred
galaxies 
\citep{2010ApJ...715L..56S},
supporting the idea of bar driving.
However, this correlation may simply mean that conditions
that favor bars also favor strong arms
\citep{2019A&A...631A..94D}.

Simulations of spirals produced by galaxy interactions
generally have grand design morphologies,
but branching of the arms is sometimes 
seen 
\citep{2015ApJ...807...73O}.
One-armed spirals, although rare, may also be linked
to galaxy interactions 
\citep{1989A&A...211...25T, 2013MNRAS.429.1051C}.
The grand design 
spirals present in
M51-like
galaxies are likely driven by the interaction 
\citep{2010MNRAS.403..625D},
but whether the spirals in all 2-armed galaxies 
were produced by interactions is a matter of debate.
Observationally, 
spiral galaxies in clusters are more likely to be 2-armed
compared to field galaxies 
\citep{2011JKAS...44..161C,
2014JKAS...47....1A,
2016MNRAS.461.3663H, 2022AJ....164..146S}.
However,
cluster spirals are also more likely
to have large bulges than field spirals 
\citep{2003MNRAS.346..601G, 2009MNRAS.394.1213W,
2022AJ....164..146S}.
and galaxies with larger bulges are more likely to be 2-armed.
When comparing galaxies with similar concentrations,
\citet{2022AJ....164..146S}
found similar arm counts in cluster
and field galaxies.
They concluded that
the primary factor that determines the number of arms
is the concentration rather than the environment.

The spiral arms in
zoom-in hydrodynamical cosmological simulations 
of Milky Way-like galaxies in Local Group-like environments 
or more isolated locations have been studied by 
\citet{2025arXiv250722793Q}
and 
\citet{2025arXiv251025848G}.
\citet{2025arXiv250722793Q}
find that 
2-armed patterns
dominate the stellar morphologies of their model galaxies.
In some cases, 
they can match a specific 2-armed pattern to a particular minor
merger, indicating that weak tidal interactions
play a role
even in relatively low density environments.
However, they
cannot link all 2-armed patterns 
to tidal interactions, 
suggesting that stochastic collapse may 
create some 2-armed morphologies.
\citet{2025arXiv250722793Q}
note that their model galaxies host multiple spiral patterns 
at once, and suggest that more than one mechanism may be needed to
explain the observed structures.

In the local Universe,
the distribution of 
spiral galaxies
in a concentration-log M* plane
shows a bend, with many more high concentration
galaxies above M* = 10$^{10}$ M$_{\sun}$ than below
\citep{2020MNRAS.493.1686L, 2022AJ....164..146S}.
This bend is also visible in plots
of
$\Sigma$$_1$, the surface brightness
in the inner kpc, vs.\ M*
\citep{2020MNRAS.493.1686L}.
The rapid increase
in concentration
above 
M* = 10$^{10}$ M$_{\sun}$ corresponds to 
a sudden increase in surface brightness. 
\citet{2020MNRAS.493.1686L}
suggest that
this bend is caused
by the growth of classical
bulges at high stellar masses.
A classical
bulge is rounder in appearance
and populated by 
stars in approximately random orbits, in contrast
to more flattened pseudo-bulges 
which show a net rotation 
\citep{2004ARA&A..42..603K,
2005MNRAS.358.1477A,
2017AA...604A..30N,
2020ApJS..247...20G}.
The bend in the 
$\Sigma$$_1$ vs.\ log M* relation
for galaxies 
has been detected out to z $\sim$ 3 
\citep{2017ApJ...840...47B}.
The bend may be a product of
the non-simultaneous growth of 
the 
bulge, the halo, and the stellar disk
due to mergers, gas accretion, star formation, feedback, and quenching
over the history of the Universe
\citep{2020MNRAS.493.1686L, 2017ApJ...840...47B, 2025MNRAS.537.3929C}.
In the local Universe,
the bend in the C-log M* diagram is particularly noticeable when the sample
is limited to 
grand design
galaxies; 
the majority of the high concentration, high mass
spiral galaxies are grand design, while
high
mass galaxies with lower concentrations
tend to be multi-armed
\citep{2024AJ....168...12S}.

Whether the local relations between
the number of arms, the concentration, and
the stellar mass of spiral galaxies extend to
higher redshifts is uncertain.
With 
Hubble Space Telescope rest-frame optical 
images, only about
10\% of galaxies at z $\sim$ 1.5 can be identified as spirals
\citep{2022MNRAS.511.1502M}, but 
the 
higher sensitivity and better resolution imaging of
JWST indicates larger percentages of spirals.
\citet{2024ApJ...968L..15K}
report a spiral fraction of about 40\% at z $\sim$ 0.75;
they conclude that the fraction drops to 30\% at z $\sim$ 3
after they correct
for resolution effects.
In another JWST study of 10$^{10}$ M$_{\sun}$ 
$\le$ M*
$<$ 10$^{11.4}$ M$_{\sun}$ galaxies,
\citet{2025A&A...700A..42E}
conclude that the fraction of spirals decreases from 55\% at z $\sim$ 1 to
20\% at z $\sim$ 2.25.  
\citet{2025A&A...700A..42E}
classify about 2/3rd of the
spirals in their sample as 2-armed, and approximately 1/3 as 3-armed.
For comparison to these numbers,
\citet{2016MNRAS.461.3663H}
use
Galaxy Zoo 2 citizen scientist classifications
\citep{2013MNRAS.435.2835W} to
estimate that 
64\% of spirals 
with M* $\ge$ 10$^{10.6}$ M$_{\sun}$
in the local Universe
are 2-armed, and 18\% are 3-armed.
\citet{2025A&A...700A..42E}
conclude that the 
fraction of 2-armed 
galaxies at z $\sim$ 1
is similar to the 
\citet{2016MNRAS.461.3663H}
local values,
but the numbers of 3-armed galaxies at high
redshift 
is
enhanced relative to local galaxies. 
\citet{2025A&A...700A..42E}
suggest that the larger gas
fractions, enhanced turbulence, 
and higher star formation rates (SFRs) at higher redshifts
favor 3-armed morphologies.  This idea is supported by
simulations by 
\citet{2024ApJ...968...86B},
which show that gas-rich
turbulent disks can produce transient 3-armed patterns in their
inner regions.

To investigate the relations between
arm number, concentration, stellar mass, and redshift, 
it would be helpful to study galaxies
in the intermediate redshift range of 0.2 $-$ 1 to connect
trends seen in the local Universe with those at higher redshifts. 
To this end, we present a first attempt to observationally
determine 
the spiral-arm-number-concentration-M* relation 
for galaxies with
0.2 $<$ z $<$ 1 using Euclid satellite 
\citep{2025A&A...697A...1E}
galaxy catalogs 
\citep{2025arXiv250315305E,
2025arXiv250315310E}.
In Section \ref{sec:data}, we describe the Euclid dataset and our sample
selection, and investigate the fraction of spirals that are 2-armed 
and 3-armed as a function of redshift.
We explore the relations between
concentration, mass, and SFR for 2-armed vs.\ 3-armed 
galaxies in Section \ref{sec:results}.
In Section \ref{sec:1arm}, we discuss the 1-armed galaxies.
In 
Section \ref{sec:discussion}, we compare with earlier studies and discuss
the implications of our results.
A summary is given in Section \ref{sec:summary}.
As in the earlier Euclid papers, we assume the Planck 2016
flat $\Lambda$CDM cosmology with H$_0$ = 67.74 km~s$^{-1}$Mpc$^{-1}$,
$\Omega$$_{\rm m}$ = 0.3089, and $\Omega$$_{\Lambda}$ = 0.6911 
\citep{2016A&A...594A..13P}.

\section{Data and Sample} \label{sec:data}

\subsection{Euclid Data and Catalogs}

The Euclid satellite 
\citep{2025A&A...697A...1E}
is currently mapping the sky
at both visible and near-infrared wavelengths.
In this paper, we will focus on images obtained with the 
Visible Imager (VIS), which uses a broadband optical 
I$_{\rm E}$ filter (0.53 $-$ 0.92 $\mu$m) and
provides a FWHM spatial resolution of 0\farcs18
and a pixel size of 0\farcs1
\citep{2025A&A...697A...2E}.
The Euclid Quick Release 1 (Q1) data 
release covers a total of 63.1 degree$^2$,
in three fields: the Euclid Deep Field North (EDF-N), 
Euclid Deep Field South (EDF-S), and Euclid Deep Field Fornax (EDF-F)
\citep{2025arXiv250315302E}.
The Euclid Q1 data release `merged' MER catalog 
\citep{2025arXiv250315305E}
contains about 29 million sources.
The MER catalog provides fluxes, 
the spatial area of the galaxy in the segmentation map 
(segmentation\_area), 
S\'ersic fits to the 
radial
light profiles,
estimates of the concentration,
and many other parameters.
The 
concentration 
is defined as 
C = 5 log$_{\rm 10}$ (r$_{80}$/r$_{20}$)
\citep{1985ApJS...59..115K, 
2014ARA&A..52..291C}.

Using both Euclid and ground-based data,
photometric redshifts and physical parameters
were derived for 
about 26 million of the MER catalog galaxies,
and 
tabulated in the PHZ catalog 
\citep{2025arXiv250315306E}.
The PHZ catalog provides two estimates of redshifts:
photometric redshifts (PHZ\_MEDIAN) from the
template-fitting {\it Phosphoros} software, and
redshifts inferred from spectral energy distribution (SED) fits for 
the physical parameters of the galaxies
obtained with the Nearest-Neighbour Photometric Redshifts
(NNPZ) software.  The PHZ catalog parameters 
PHZ\_PP\_MEDIAN\_Z and 
PHZ\_PP\_MODE\_Z are the median and mode of the NNPZ redshifts.
The NNPZ SED-fitting routine also provides estimates of the stellar mass
and SFR.
\citet{2025arXiv250315306E}
compared the PHZ redshifts with spectroscopic
redshifts, and concluded that the accuracy of the photometric
redshifts,
defined
as the median $|$z$_{phot}$ - z$_{spec}$$|$, 
is about 2\%, while the precision, as measured
by the normalized median absolute deviation, is 0.03 for
I$_{\rm E}$ $<$ 23.   
They find a outlier fraction of 10\% in the north
and 16\% in the south, where outliers are defined as galaxies
for which the photometric redshift is off by more than 0.15.

\citet{2025arXiv250315310E}
used the Zoobot galaxy morphology foundation model to 
derive morphological
parameters for a subset of 380,111 of the
Euclid 
galaxies in the MER catalog. 
The Zoobot software 
mimicked Galaxy Zoo 2 citizen scientist responses
to a set of questions about each galaxy.  These questions
include whether the galaxy is smooth or featured, whether it is
an edge-on disk, and whether a spiral pattern is visible. 
If those questions are answered positively, then citizen scientists
were asked how many arms were present.  
The Zoobot catalog provides synthetic `vote fractions' for
each galaxy in the sample, mimicking
the expected responses that a set of random citizen scientist
volunteers would have selected for each galaxy.
These vote fractions were not corrected for redshift classification bias,
so are equivalent to the `raw' vote fractions used
by 
\citet{2016MNRAS.461.3663H}.

The Zoobot was 
pre-trained using Euclidized Hubble Space Telescope
images, 
then trained 
using 
Galaxy Zoo
volunteer annotations of preliminary Euclid Wide
Survey images.  This Euclid training set was selected based
on the criteria segmentation\_area $>$ 1200 pixels
OR I$_{\rm E}$ $<$ 20.5 AND segmentation\_area $>$ 200
pixels.   In the production of the morphological
catalog, the sample was extended to
fainter (I$_{\rm E}$ $<$ 23)
galaxies, but with 
a segmentation\_area limit of 
$>$ 700.   This means that the software 
was applied to some galaxies that were fainter 
than the original training set.
To exclude artifacts, 
\citet{2025arXiv250315310E}
limited their catalog to galaxies
with the MER parameter vis\_det set to 1 (i.e., the object
was detected
in the visible), and 
the spurious\_prob parameter set to $<$ 0.2.

\subsection{Sample Selection}

We started with the 
\citet{2025arXiv250315310E}
Euclid morphology
catalog
of 380,111 galaxies
\citep{walmsley_2025_15106473}.
We cross-correlated with the Dark Energy Spectroscopic Explorer (DESI)
Data Release 1
spectroscopic redshift catalog 
\citep{2025arXiv250314745D}
using a 2$''$
offset, and found 18,673 matches. 
When available, we used the DESI spectroscopic
redshift in the following analysis; otherwise, we used the 
NNPZ photometric redshift PHZ\_PP\_MEDIAN\_REDSHIFT.
Following 
\citet{2025arXiv250315309E},
we eliminated objects with axis ratio
$\le$ 0.05
(i.e., sersic\_sersic\_vis\_axis\_ratio parameter $\le$ 0.05), 
since those are likely artifacts, and 
restricted the sample to
galaxies with reliable physical parameters in the PHZ
catalog (phys\_param\_flags = 0).  As in 
\citet{2025arXiv250315309E},
we eliminated galaxies with large differences between
the two photometric determinations of redshift, only retaining
galaxies with 
$|$ PHZ\_PP\_MEDIAN\_Z - PHZ\_MEDIAN $|$ $<$ 0.2 
and 
$|$ PHZ\_PP\_MEDIAN\_Z - PHZ\_PP\_MODE\_Z $|$  $<$ 0.2 unless
a spectroscopic redshift is available.
To remove spurious and compact sources, we required DET\_QUALITY\_FLAG $<$ 4
and MUMAX\_MINUS\_MAG $>$ -2.6, as in 
\citet{2025arXiv250315306E}.
We also excluded galaxies with PHZ specific SFRs (sSFRs) $>$ 10$^{-8.2}$ yr$^{-1}$, as their
parameters may be unreliable 
\citep{2025arXiv250315306E}.
In our final analysis,
we only included galaxies with
both concentration and stellar mass
available from the PHZ catalog, and
S\'ersic index from the MER catalog.
We also eliminated
objects with unreliable physical parameters
by requiring the stellar mass to be less than 10$^{12}$ M$_{\sun}$ and
the concentration to be less than 20.

The Euclid morphological catalog 
\citep{2025arXiv250315310E}
only
provides Zoobot spiral arm count vote fractions 
for galaxies 
with p$_{featured}$ $\times$ p$_{notedge-on}$ $\times$ p$_{spiral}$ $>$ 0.5,
where 
p$_{featured}$, p$_{notedge-on}$, and p$_{spiral}$ 
are the Zoobot's estimated vote fractions for the 
questions about smoothness vs.\ features/disk, non-edge-on disks,
and the presence of a spiral pattern, respectively.
We limited our sample to spiral galaxies with effective radii
(half-light radii) greater than or equal to 0\farcs5, as determined
by the S\'ersic-fitting routine (parameter sersic\_sersic\_vis\_radius).
This radius is 3$\times$ the FWHM resolution, sufficient
for reasonable morphological study.  This requirement only
eliminates a few galaxies that meet the other criteria.
Limiting the initial sample to 0.15 $\le$ z $\le$ 1.2,
a total of 32,974 galaxies meet the above criteria;
11,514 of these are in the EDF-N field, 
15,468 in the EDF-S field, and 
5992 in the EDF-F field.
Of the 32,974 galaxies in our initial sample of spirals, 2172 have DESI
spectroscopic redshifts.
The full set of criteria used to select our initial
sample of spiral galaxies is given
in Table 
\ref{tab:selection}.

\begin{deluxetable*}{cccc}
\tablecaption{Galaxy Sample Selection Criteria\label{tab:selection}}
\tablewidth{0pt}
\tablehead{
\colhead{Catalogue} & \colhead{Parameter(s) and Criteria} 
& \colhead{Source}
& \colhead{Notes}
}
\startdata
MER & segmentation\_area $>$ 1200 pix & \citet{2025arXiv250315310E} \\
 &  {\bf OR} & \\
MER & (I$_{\rm E}$ $<$ 20.5 AND segmentation\_area $>$ 200 pix) & \\
MER & vis\_det = 1 & \citet{2025arXiv250315310E} \\
MER & spurious\_prob $<$ 0.2 & \citet{2025arXiv250315310E}\\
PHZ  &  phz\_param\_flags = 0 &  \citet{2025arXiv250315309E} \\
PHZ  & $|$ PHZ\_PP\_MEDIAN\_Z - PHZ\_MEDIAN $|$ $<$ 0.2 & \citet{2025arXiv250315309E} & 1 \\ 
PHZ  & $|$ PHZ\_PP\_MEDIAN\_Z - PHZ\_PP\_MODE\_Z $|$  $<$ 0.2 & \citet{2025arXiv250315309E} & 1 \\ 
MER & spurious\_flag = 0 & \citet{2025arXiv250315309E} \\
MER & sersic\_sersic\_vis\_index $<$ 5.45 & \citet{2025arXiv250315309E} \\
MER & sersic\_sersic\_vis\_axis\_ratio $>=$ 0.05 & \citet{2025arXiv250315309E} \\
MER & mumax\_min\_mag\_x/y $\ge$ $-$2.6 & 
\citet{2025arXiv250315306E}  \\
MER & det\_quality\_flag $<$ 4 & 
\citet{2025arXiv250315306E}
\\
PHZ & phz\_flag = 0 & 
\citet{2025arXiv250315306E}
\\
PHZ & log sSFR $<$ $-$8.2 & 
\citet{2025arXiv250315306E}
\\
PHZ & phz\_pp\_median\_stellarmass available & This work \\
PHZ & phz\_pp\_median\_sfr  available & This work \\
PHZ & phz\_pp\_median\_stellarmass $<$ 12 &  This work \\
MER & concentration $<$ 20 & This work \\
MER & sersic\_sersic\_vis\_radius $\ge$ 0.5 & This work \\
MER & p$_{featured}$ $\times$ p$_{notedge-on}$ $\times$ p$_{spiral}$ $>$ 0.5 & 
\citet{2025arXiv250315310E} \\
   &  0.15 $\le$ z $\le$ 1.2 &  Our initial spiral sample\\
   &  log (M*/M$_{\sun}$) $\ge$ 10.6 &  Our mass-limited spiral sample\\
morph & spiral-arm-count\_2\_fraction $\ge$ 0.8 & Our highly reliable 2-armed sample\\ 
morph & spiral-arm-count\_3\_fraction $\ge$ 0.5 & Our highly reliable 3-armed sample\\ 
  & 8.5 $\le$ log(M*/M$_{\sun}$) $<$ 12 & Our highly reliable samples\\
\enddata
\tablenotetext{1}{Unless spectroscopic redshift available from DESI.}
\end{deluxetable*}

\subsection{The Fraction of Spirals with Different Arm Counts, as a Function of Redshift}

We initially assigned each galaxy in the spiral sample
to one of the following six classes:
1-armed, 2-armed, 3-armed, 4-armed, 5+-armed, or `can't tell the number of arms', by
placing them in the class for which the Zoobot gave the highest
vote fraction. 
To determine how the fraction of spiral galaxies of each class
changes with redshift, we need to 
take into account completeness as a function of mass, which depends
upon redshift.
As we show below (Section \ref{sec:mass_vs_redshift}),
at z = 0.2 the sample is about 90\% complete to log (M*/M$_{\sun}$) = 10.0.
By a redshift of z = 0.5, the sample is only 50\% complete
at log (M*/M$_{\sun}$) = 10.6.
In the following discussion of the fraction of spirals
in the different classes as
a function of redshift, we 
only include galaxies with 
log (M*/M$_{\sun}$) $\ge$ 10.6.
We call this the mass-limited spiral sample.
Note, however, that
at redshifts greater than 0.5 this mass-limited sample is somewhat
incomplete
even with a 
log (M*/M$_{\sun}$) $\ge$ 10.6 limit.
The total number of galaxies in the mass-limited sample
is 10,820 galaxies.

\begin{figure*}[htp]
\plottwo{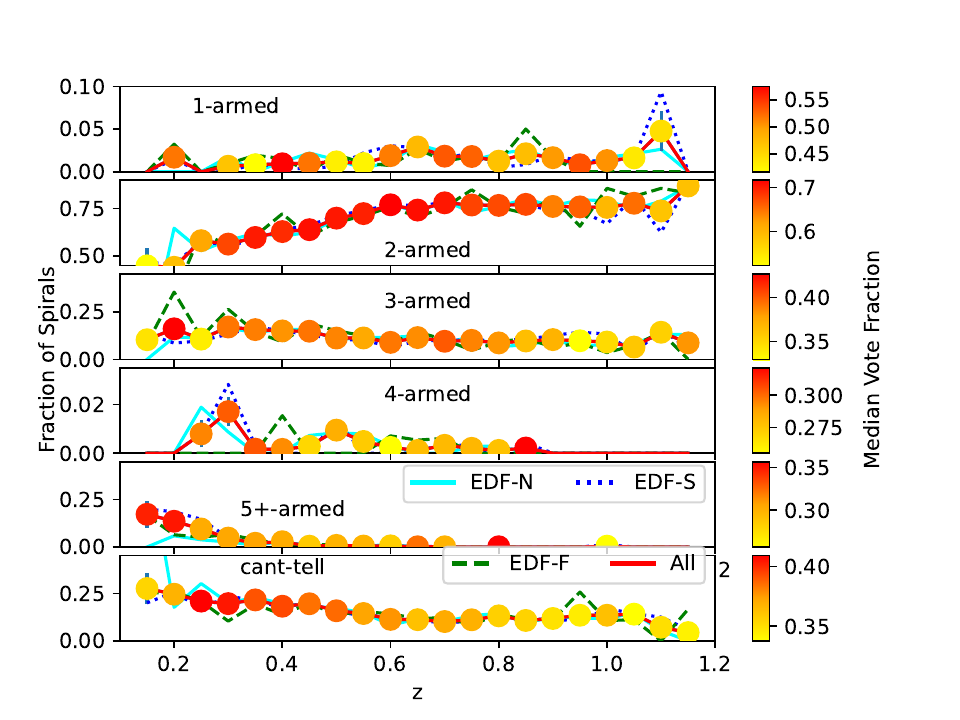}{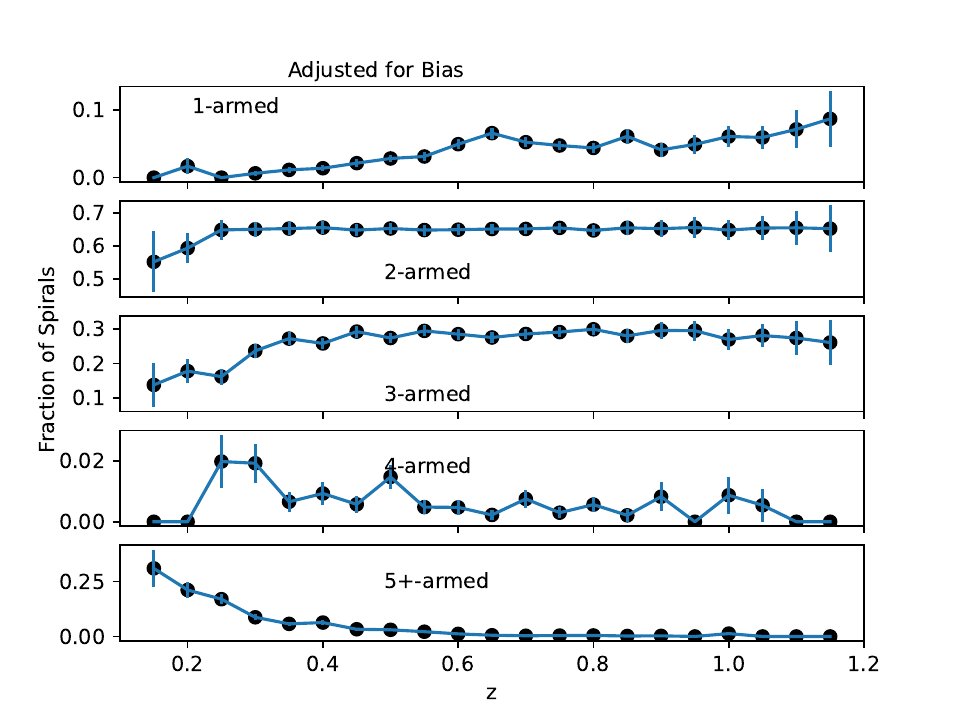}
\caption{
Left:
The fraction of M* $\ge$ 10$^{10.6}$ M$_{\sun}$
spiral galaxies that are placed in the
various 
arm count categories,
as a function of redshift.
These fractions were
calculated using the raw vote fractions from the Zoobot,
and include the `can't tell' class.
The solid red line and colored points are for the full set of data;
the cyan solid line, dotted blue line, and green dashed line 
are for EDF-N, EDF-S, and EDF-F,
respectively.
The datapoints are color-coded from yellow to red based on
the mean vote fraction for that datapoint, as shown by the individual
colorbars to the right.
Right: 
The fraction of M* $\ge$ 10$^{10.6}$ M$_{\sun}$
galaxies that are placed in the
arm count categories,
after correction for bias.  
The `can't-tell' galaxies are not included in the
statistics shown on the right.
For these bias correction,
the fraction of galaxies
identified as 2-armed is fixed at 65\%,
and the 2-armed vote fractions are scaled down 
from their raw values
accordingly.
\label{fig:raw_fractions}}
\end{figure*}

For this mass-limited spiral sample,
the fraction of
the spiral 
galaxies in each of these classes (including the `can't-tell' class)
is plotted as a function
of redshift in Figure \ref{fig:raw_fractions}.
To construct this figure,
we divided the galaxies into redshift bins of width $\Delta$z = 0.05 
for 0.15 $<$ z $<$ 1.2.  
At all redshifts, 
the majority of the galaxies 
are placed in the 2-armed category. 
Ignoring the lowest redshift bin as having poor statistics and 
averaging over the second and third redshift bins
(0.175 $\le$ z $<$ 0.275),
the percentage of galaxies with log (M*/M$_{\sun}$) $\ge$ 10.6 classified
by the software as 1-armed/2-armed/3-armed/4-armed/5+-armed/cant-tell
are 0.5 $\pm$ 0.4\%, 54 $\pm$ 3\%, 
12 $\pm$ 2\%, 0.05 $\pm$ 0.4\%, 11 $\pm$ 2\%, and 22 $\pm$ 2\%,
respectively.
The Zoobot places almost 1/4th of the galaxies at z $\sim$ 0.2 in
the `can't-tell' class.  These 'can't-tell' galaxies may be a combination of 
flocculent galaxies and multi-armed galaxies for which the arm counts
are unclear.  The fraction of galaxies listed as `can't-tell' drops off
with redshift, perhaps because at higher redshifts the software is
no longer able to discern a spiral pattern, so the galaxies are
not included in the spiral sample.

In Figure
\ref{fig:raw_fractions}, we provide curves for
the EDF-N, EDF-S, and EDF-F fields separately, in addition to 
values for the sample as a whole.
Around z = 0.2, 
there is considerable scatter from field to field.
At z = 0.2, the EDF-N field has a much higher fraction of 2-armed
galaxies compared to the other fields.
A comparison of 
the Euclid photometric redshifts 
with 
published spectroscopic redshifts showed that 
in the EDF-N field, the outliers cover a spread in redshift, but
in the EDF-S and EDF-F fields,
a large fraction of the outliers 
were
at z$_{phot}$ = 0.2 $-$ 0.3 and z$_{spec}$ = 0.5 $-$ 0.8 
\citep{2025arXiv250315306E}.
This means that the z = 0.2 statistics are less reliable for 
EDF-S and EDF-F than for EDF-N.
These redshift errors may contribute to the scatter seen 
at z $\sim$ 0.2
in Figure \ref{fig:raw_fractions}. 
Since about 3/4th of the sample galaxies
are in the southern and Fornax fields,
the overall statistics at z = 0.2 may be skewed by these outliers.

In the statistics given below,
to be consistent with 
\citet{2016MNRAS.461.3663H}
we exclude galaxies identified as
`can't-tell' from the quoted fractions.
Excluding 
`can't-tell' galaxies,
65 $\pm$ 2\% of the 
0.175 $\le$ z $<$ 0.275
spirals 
are placed in the 2-armed class and 15 $\pm$ 2\%
in the 3-armed class.
These percentages are uncertain, 
however, because of the field-to-field scatter
at z = 0.2
noted above. 
Between z = 0.20 and 0.25
the percentage of 2-armed galaxies in the full sample increases from
57 $\pm$ 5\% to 69 $\pm$ 3\%, 
but in the EDF-N field, 
the percentage drops from 76 $\pm$ 10\% to 66 $\pm$ 7\%.

Over a larger redshift range, there is a clear increase with redshift
in the 
fraction of galaxies classified as 2-armed.
The percentage 
of 2-armed galaxies increases from $\sim$65\% at low redshifts to 84\%
at z $\sim$ 0.6.
The fraction of spirals placed in the 3-armed category
is about 15 - 20\% from 0.2 $\le$ z $<$ 0.4, then 
gradually drops to about 15\% at z $\sim$ 0.6 $-$ 0.8.
The percent of 4-armed galaxies is about 1 $-$ 2\% at 
z $\sim$ 0.2 and drops to $\sim$ 0.5\% at z $\sim$ 0.6.
The fraction of galaxies classified as 
5+-armed is substantial 
at
z $\sim$ 0.2 
(21\%), but drops quickly at higher redshifts, falling to about 6\% at
z = 0.4 and 1\% at z = 0.6.
In contrast to the 3-armed, 4-armed, and 5+-armed fractions, the fraction of
1-armed galaxies increases slightly with redshift, though it remains small.
It is 0.5 $\pm$ 0.4\% at 0.175 $\le$ z $<$ 0.275, 1.1 $\pm$ 0.2\% at 
0.4 $\le$ z $\le$ 0.6, and 1.6 $\pm$ 0.3\% at 0.8 $\le$ z $<$ 1.

The datapoints in the left panel of Figure 
\ref{fig:raw_fractions} are color-coded based on the median raw vote fraction
for galaxies in that class.  For galaxies that we classify as 2-armed,
the median vote fraction for two arms is typically 0.6 $-$ 0.7, i.e.,
a majority of the votes.
In contrast, for the galaxies in the other classes, the median vote
fraction in their selected class is typically less than 0.5, meaning
that, although a plurality of the votes placed them in their class,
it was not the majority.   

The redshift trends in the fractions of galaxies in each class 
are likely caused at least in part by
classification bias.  At higher redshifts, viewers may 
undercount the number of arms, causing the dramatic
drop in 5+-armed systems between z = 0.2 and z = 0.6.
For more nearby galaxies, one might be able to see further out
in the disk where branching of arms may occur, giving a higher arm count.
More generally, 
multi-armed galaxies may be
mis-identified
as 2-armed at higher redshifts.
Furthermore, at higher redshifts
the spiral patterns in multi-armed and flocculent
galaxies may be hard to see, thus
such galaxies are more likely to be omitted from
the spiral dataset at higher redshifts.

For comparison to these numbers, 
\citet{2016MNRAS.461.3663H}
found 
bias-corrected 
percentages 
(excluding `can't-tell' galaxies)
of 1-armed/2-armed/3-armed/4-armed/5+-armed
spiral galaxies 
of
5\%/64\%/18\%/6\%/7\%
for galaxies with log (M*/M$_{\sun}$) $\ge$ 10.6
in the 0.03 $<$ z $<$ 0.085 range.
Their percentages of 2-armed and 3-armed agree well with ours
at 0.2 $<$ z $<$ 0.4, showing good consistency 
between the
Euclid Zoobot values 
and the SDSS Galaxy Zoo 2 
data.
However,
the Euclid Zoobot finds a higher fraction of 5+armed galaxies
and a lower fraction of 1-armed and 4-armed galaxies
at z = 0.2 
than 
\citet{2016MNRAS.461.3663H}
found for SDSS galaxies at z = 0.
The difference in the fraction of
5+-armed galaxies may be a consequence of the excellent
spatial resolution of Euclid.
At a redshift of 0.2 the Euclid spatial resolution
of 
0\farcs16 
corresponds to a spatial scale of 0.53 kpc,
while at a redshift of 0.03, the lower edge of the SDSS
range studied by 
\citet{2016MNRAS.461.3663H},
the SDSS spatial
resolution of 
1\farcs3 
corresponds to a spatial scale of 0.79 kpc.
In Section \ref{sec:disc_2arm_vs_3arm},
we compare our Euclid statistics with other
studies besides 
\citet{2016MNRAS.461.3663H}.

The strong trends 
with redshift indicate that classifier bias is present in
the raw Euclid Zoobot vote fractions.   
As a first attempt to disentangle the effects of observer bias
and galaxy evolution in 
the observed trends,
we will
make a simple assumption.   
JWST observations find that about 
$\sim$60\% of spirals 
with 10$^{10}$ M$_{\sun}$ $\le$ M* $<$ 10$^{11.4}$ M$_{\sun}$ are 2-armed 
at z = 1 
\citep{2025A&A...700A..42E},
similar 
to the 
\citet{2016MNRAS.461.3663H}
SDSS percentage and also similar to 
the 65\% we find from Euclid at z $\sim$ 0.2.  
We will therefore naively assume that the
fraction of 2-armed galaxies remains constant 
with redshift 
at 65\% for 0.2 $<$ z $<$ 1.
This is an over-simplistic assumption, 
since the spiral structures
of galaxies may 
evolve significantly between z = 1 and z = 0.2.
However, given the apparent agreement for 2-armed galaxies 
between SDSS, JWST, and z $\sim$ 0.2 Euclid, we willl 
assume that the fraction of 2-armed spirals is constant,
and 
then investigate how the fractions
for the other classes change.

At each redshift bin, for each galaxy we multiply the Zoobot vote fractions
for 2-arms by a scale factor less than 1, to
account for the apparent observer bias.   We empirically determine
what this scale factor is at each redshift by requiring the 
fraction of 2-armed spirals to remain fixed at 65\%.  
To be consistent with 
\citet{2016MNRAS.461.3663H},
we have excluded galaxies
in the `can't-tell' class from the bias correction.
The `can't-tell' galaxies were excluded from the sample, and the
bias correction was done on the remaining galaxies.
The application of
this scale factor will move some of the galaxies previously
listed as 2-armed to other categories.
For a given redshift, we use the same scale factor 
for galaxies of all masses,
rather than binning the galaxies
further by mass and size as done by 
\citet{2013MNRAS.435.2835W}.
The corrected vote fractions as a function of redshift 
are displayed
in the right panel of Figure 
\ref{fig:raw_fractions}.
In contrast to the left panel of Figure 
\ref{fig:raw_fractions}, the right panel 
excludes
the can't-tell galaxies.

Since typical vote fractions for 3-armed galaxies are usually
higher than for 4-armed or 5+-armed systems (left panel, Figure \ref{fig:raw_fractions}),
when the vote fractions for 2-armed are lowered in
the correction process, 
galaxies are more likely moved from the 2-arm class to the 3-arm class.
Therefore, by requiring the 2-arm fraction to be fixed with redshift,
we increase the 3-arm fraction at higher redshift.  In the corrected
plots, the 3-arm fraction rises quickly from 20\% at z = 0.2
to 30\% at z = 0.3, and then remains constant at about 30\% to z = 1.
This rapid rise in the corrected fraction of 3-armed
galaxies between z = 0.2 and z = 0.3 is probably artificial,
perhaps
caused by the systematic problems with the photometric redshifts 
at z = 0.2 
which cause the field-to-field variations.
Many of the galaxies reclassified
as 3-armed by the bias correction may actually be 4-armed or 5+-armed.

Another possible artifact in the corrected-for-bias
plots is the increase in 
the fraction of galaxies classified as 1-arm between z = 0.2
and z = 1.   As shown in the 
left panel of Figure \ref{fig:raw_fractions},
the median vote fractions for galaxies selected as 1-armed
are relatively high compared to 
the 4-armed and 5+-arm classes.  Although the 
fraction of 
galaxies selected as 1-armed is low, less than
3\%, galaxies that are selected as 1-armed tend to have relatively
high vote fractions for 1-armedness.
This means that when
galaxies are not moved to the 3-armed class by the bias correction, they 
tend to be moved to the 1-arm class.  

As a rough estimate of the uncertainties in the corrected
fractions of galaxies of each arm count, 
we calculated bias-corrected fractions assuming
the fraction of 2-armed galaxies is constant at 60\% or 70\% rather than
the nominal 65\%.   These changes caused the corrected fractions
for 3-armed galaxies to increase to $\sim$35\% or decrease to $\sim$25\%,
respectively.

The above discussion illustrates the large uncertainties in correcting
for redshift classification bias in the Euclid database.  
Even with these
large uncertainties, however,
it is intriguing that our bias-corrected 3-arm fractions of 30\%
at z = 1 agree well with the JWST 
fractions at z = 1 reported by 
\citet{2025A&A...700A..42E}.
This result is quite uncertain, however, and requires further
morphological analysis of the structure of the spiral arms, 
including an exploration
of possible branching in the arms. 
A more detailed study of redshift classification bias in
the Euclid Zoobot database also requires improved photometric redshifts,
and is beyond the scope of this study.

\subsection{Highly Reliable Samples of 2-Armed and 3-Armed Galaxies}

We now further filter our galaxy sets to obtain
the purest and most reliable
samples of 2-armed and 3-armed galaxies possible.  Our ultimate
goal is 
to investigate physical differences between galaxies with
different numbers of arms, thus we would like to minimize
the numbers of misclassifications.
For our final 'highly reliable' sample of 2-armed galaxies,
we use the 
subset of galaxies 
with raw vote fraction for
2-arms $\ge$ 0.8.  
For our final sample of 'highly reliable' 3-armed galaxies, we limit our
sample to 
galaxies 
with raw vote fractions for 3-arms $\ge$ 0.5.
The 2-armed galaxy
sample is therefore a purer sample than the 3-armed sample.
The less strict selection criteria used for the 3-armed 
sample 
is necessitated by the need for a reasonable sample size.
As we discuss
in Sections 
\ref{sec:C_vs_mass}
and
\ref{sec:main_seq},
comparisons of samples of equal purity do not change
our conclusions.
For both sets of galaxies,
we extend the mass range down to log (M*/M$_{\sun}$) = 8.5
to study trends with M*.

The standard Euclid catalog
pipeline background subtraction gives less reliable photometry
for large angular size nearby galaxies 
\citep{2025arXiv250315335M}.
Furthermore, the Euclid photometric redshift pipeline
was designed for galaxies
with z $\ge$ 0.2 
\citep{2020A&A...644A..31E}.
This means that the 
photometric
redshifts may be less reliable 
at low
z.
We therefore limit our sample to z $>$ 0.2
in the following
analysis. 
We limit our sample to z $\le$ 1 because of angular size constraints.
This gives us 6284 2-armed and 262 3-armed galaxies in our 
highly reliable subsets.
These samples of 2-armed and 3-armed galaxies are compared and contrasted in
Section
\ref{sec:results}.

Likely some of the galaxies classified 
as 2-armed or 3-armed have 
fainter arms and branches, and in a more detailed study 
may be catalogued
as 4-armed or multi-armed.
In the current study, we assume that the 
majority of the galaxies in our 2-armed sample
have spiral patterns that 
are dominated by a 2-armed structure, although there may be 
branches and filaments.
Likewise, we assume that the majority of the galaxies we have
included in our 3-armed sample fall into 
the general category of multi-armed
galaxies.   
\citet{2024AJ....168...12S}
investigated Galaxy Zoo 2 (GZ2) arm count 
statistics 
for galaxies identified by other methods
as either grand design or multi-armed. 
They found that galaxies in the grand design
sample generally had high GZ2 vote fractions as 2-armed.
For the multi-armed galaxies in their sample,
when they summed the Galaxy Zoo 2 vote fractions for 3-armed, 4-armed,
5+ armed, or `cannot count arms', the vote fractions were generally
greater than 50\%.   
These statistics support our association of our highly reliable
2-armed galaxies with the arm class of grand design galaxies,
and our assumption that our highly reliable 3-armed galaxies
are mostly multi-armed galaxies.

We also construct a sample of `highly reliable' 1-armed galaxies,
using a vote fraction for 1-arms
of $\ge$ 0.5.   A total of 188 galaxies with 0.2 $<$ z $\le$ 1 
meet these criteria.  These galaxies are discussed further in Section
\ref{sec:1arm}.
No galaxies 
have 
vote fractions for 4-arms or 5+-arms greater than 0.5, so we do not 
construct 'highly reliable' samples of 4-armed or 5+-armed galaxies. 

\section{2-armed vs.\ 3-armed Galaxies} \label{sec:results}

\subsection{Magnitude and Redshift Limits}

\begin{figure*}[htp]
\includegraphics[width=8.0cm,trim={0.8cm 0cm 6cm 10cm},clip=true]{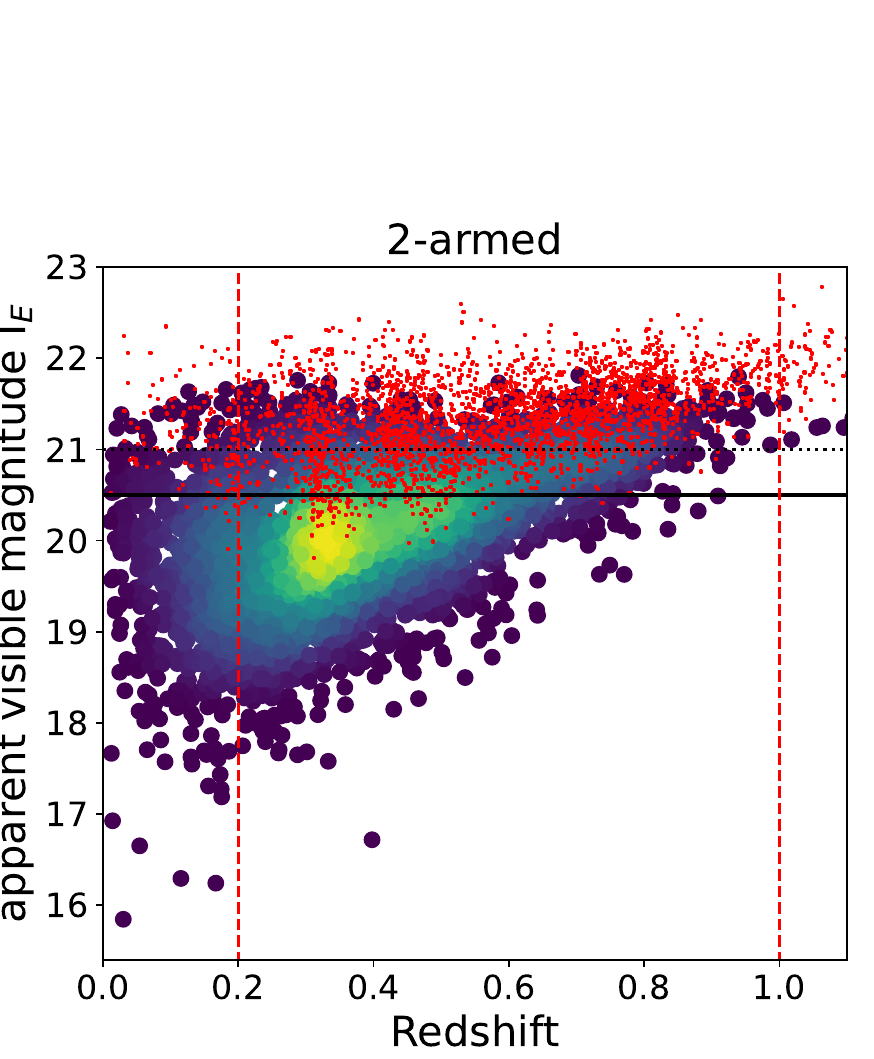}%
\includegraphics[width=9.05cm,trim={0.8cm 0cm 4cm 10cm},clip=true]{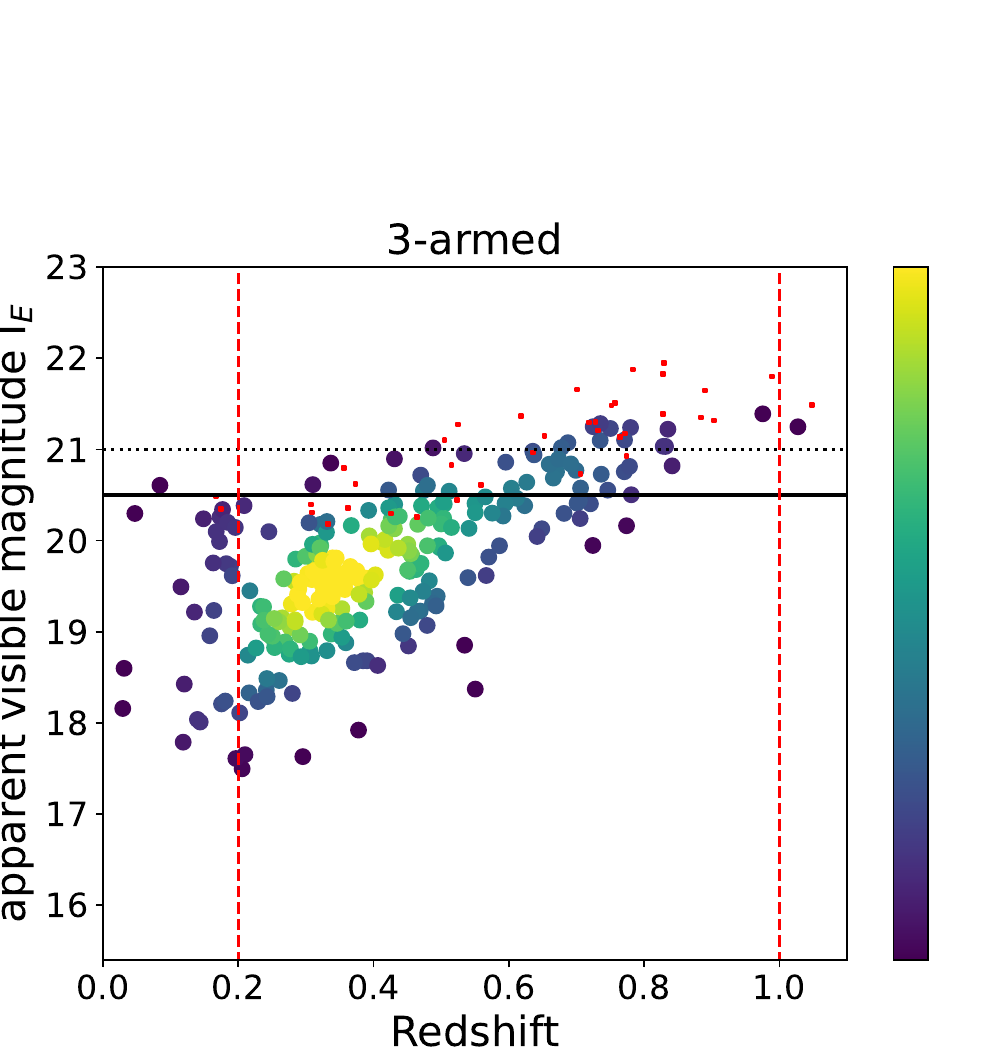}
\caption{The apparent visible magnitude I$_{\rm E}$ vs.\ redshift 
for our samples of highly-reliable 2-armed galaxies (left) and
3-armed galaxies (right).  
The red dots represent galaxies with segmentation\_area $<$ 1200 pixels.
For galaxies 
with 
larger segmentation\_areas, 
the colorbar gives 
the relative number of galaxies per
area on the plot as shown by the colorbar on the right.
The horizontal
solid black line at I$_{\rm E}$ = 20.5 is the magnitude limit
of the Zoobot training sample for segmentation\_area $<$ 1200 galaxies.
The horizontal dotted black line is our estimate of the approximate completeness
limit
of our samples.
As discussed in the text, we limit our final sets of
highly reliable
2-armed and 3-armed galaxies to galaxies with 0.2 $\le$ z $\le$ 1,
as marked by the vertical red dashed lines.
In this plot, we include galaxies outside that redshift
range for comparison purposes.
\label{fig:magredshift}}
\end{figure*}

In Figure \ref{fig:magredshift}, we plot I$_{\rm E}$ vs.\ redshift for
our samples of highly reliable 2-armed (left) and 3-armed (right) galaxies.  
In these plots,
we distinguish
between galaxies with segmentation\_area $<$ 1200 pixels and 
larger galaxies.  For a given redshift, galaxies with smaller angular sizes
tend to be fainter, as expected.  
In both panels, the highest density of points
occurs at a redshift of $\sim$ 0.3, but the peak of the distribution
for the 2-armed galaxies is at a higher magnitude of $\sim$19.9
compared to $\sim$19.5 for the 3-armed galaxies (i.e., at z $\sim$ 0.3,
the distribution of galaxies peaks at a lower luminosity for 2-armed
galaxies).

In Figure \ref{fig:magredshift}, 
we overlay (solid horizontal
line) the I$_{\rm E}$ = 20.5 limit of the 
Zoobot training sample for galaxies with segmentation\_area $>$ 200 pixels.
Most of the segmentation\_area $<$ 1200 galaxies are fainter than this
limit.
Since the Zoobot software was not trained on 
small angular size galaxies fainter than this cut-off, 
classifications of fainter galaxies may be less reliable.
For the 2-armed galaxy sample, the distribution of
large angular size galaxies extends to 
about I$_{\rm E}$ = 21.5.  

Because of the filtering process described above, our sample is not 
complete.  The completeness
of the sample varies in a complicated way with redshift, morphology,
angular size, and location in the sky.
For example, because of differences in the available ground-based data,
the galaxies in the EDF-N field with
the largest redshift uncertainties 
tend to lie between 0.45 $<$ z $<$ 0.8, while in the other fields 
galaxies with highly uncertain
redshifts have a larger range of redshift
\citep{2025arXiv250315306E}.
Although the 
\citet{2025arXiv250315310E}
morphological catalog includes galaxies
to I$_{\rm E}$ = 23, 
based on Figure 
\ref{fig:magredshift}
we 
conservatively 
estimate an
approximate 
completeness limit for our sample 
of about I$_{\rm E}$ = 21 (dotted horizontal
line).  

\subsection{Stellar Mass vs.\ Redshift} \label{sec:mass_vs_redshift}

\begin{figure*}[htp]
\includegraphics[width=9.05cm,trim={0.0cm 0cm 4cm 10cm},clip=true]{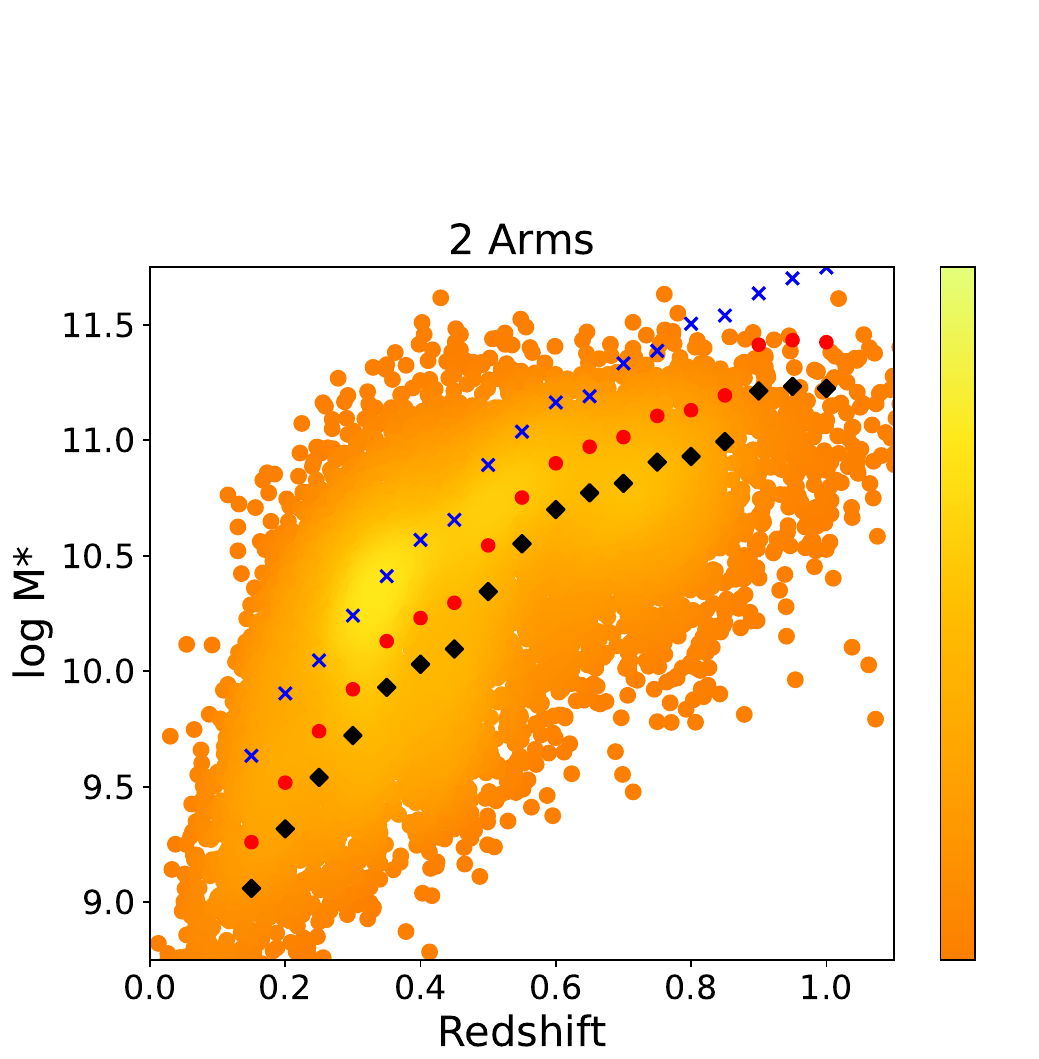}
\includegraphics[width=9.05cm,trim={0.0cm 0cm 4cm 10cm},clip=true]{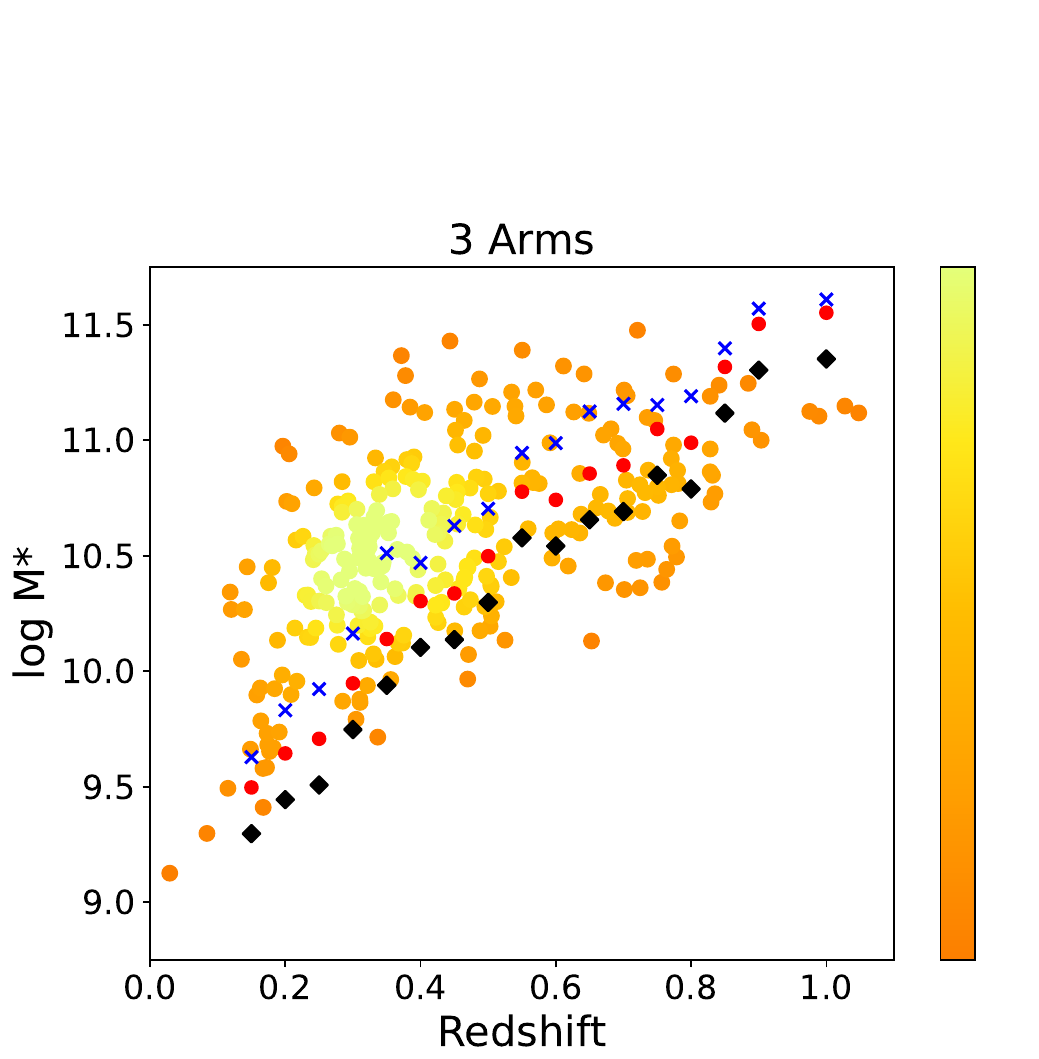}
\caption{The log M* vs.\ redshift plot for our highly
reliable samples of 2-armed (left) and
3-armed galaxies (right).  The color scale represents the relative number of galaxies per
area on the plot, as shown by the colorbar on the right.  
The 50\% completeness limit assuming a magnitude limit of I$_{\rm E}$ = 21
is plotted as black filled diamonds.  
The 50\% and 90\% completeness limits obtained using the stricter limit of 
I$_{\rm E}$ = 20.5
are plotted as red filled circles and blue crosses, respectively.
\label{fig:massredshift}}
\end{figure*}

In Figure \ref{fig:massredshift}, we plot stellar mass vs.\ redshift
for our highly reliable samples of 2-armed vs.\ 3-armed galaxies.
As expected, there is a trend in these plots,
with larger redshift galaxies tending to
be more massive.  
This result is in part
because of incompleteness, with lower mass
galaxies being missed at higher redshift.
The highest density of datapoints in these plots for both samples
of galaxies
occurs at around z = 0.3, but at a higher stellar mass (log (M*/M$_\sun$) $\sim$ 10.5) for
the 3-armed galaxies compared to the 2-armed (log (M*/M$_{\sun}$) $\sim$ 10.4).

To calculate 
completeness limits as a function of stellar mass and redshift, we used the 
method of 
Pozzetti et al. (2010): for each galaxy in the sample,
we calculated the limiting stellar
mass the galaxy would have if its apparent magnitude
were equal to the limiting magnitude of the survey.
This distribution was then used to calculate the 50\% and 90\%
completeness limit.  This method implicitly takes into account the
mass/light ratios of the sample galaxies.  We find similar
completeness limits for the 2-armed and 3-armed galaxy samples
at a given redshift
(within 0.2 dex).

In Figure \ref{fig:massredshift}, 
at each redshift
we mark the 50\% and 90\% completeness 
limit
as a function of stellar mass 
assuming our conservative magnitude
limit of I$_{\rm E}$ $<$ 20.5.
We also mark the 50\% completeness mass limit for the more liberal
magnitude limit of 
I$_{\rm E}$ $<$ 21.

\subsection{Concentration vs. Stellar Mass} \label{sec:C_vs_mass}

\begin{figure*}[htp]
\includegraphics[width=16.0cm,trim={0.2cm 0cm 6cm 15cm},clip=true]{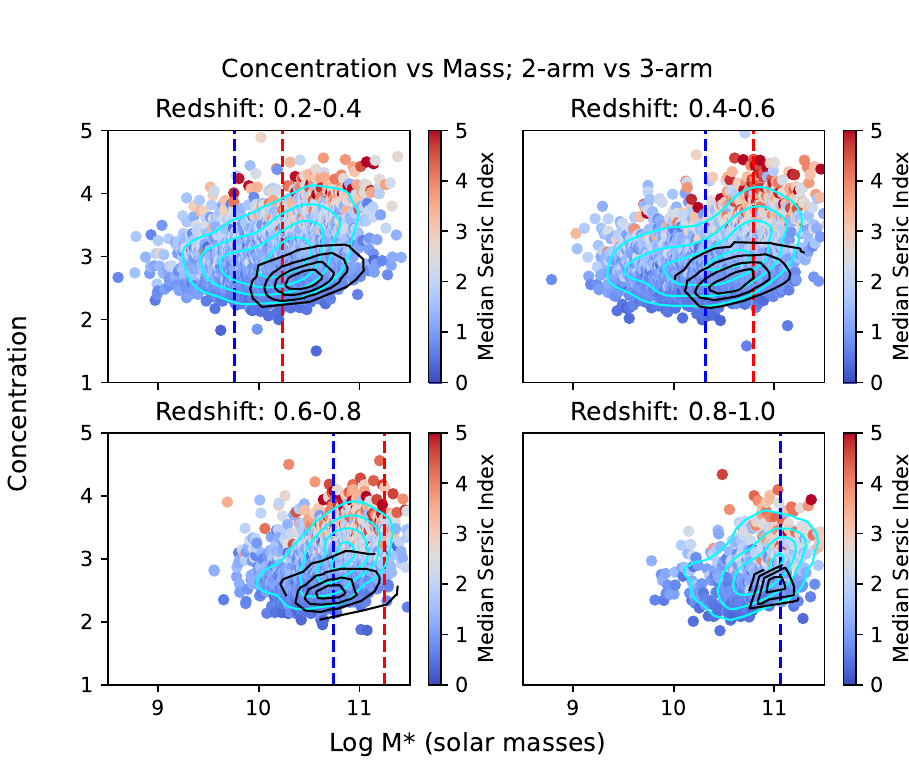}
\caption{Concentration vs. log M* for four redshift ranges,
for galaxies in our samples of highly reliable 
2-armed 
and 3-armed galaxies.
The contours give relative numbers of galaxies per area on the plot,
for
2-armed 
galaxies
(cyan contours) and 3-armed galaxies (black contours).
Concentration is defined as 
C = 5 log$_{\rm 10}$ (r$_{80}$/r$_{20}$)
\citep{1985ApJS...59..115K, 
2014ARA&A..52..291C},
where r$_{80}$ and 
r$_{20}$, respectively, are the radii containing 80\% and 20\% of the total 
Euclid visible light flux.
The background color scale gives the median S\'ersic index.
The blue dashed vertical lines represent the
50\% completeness limit assuming a magnitude limit of 
I$_{\rm E}$ = 21,
while the red dashed lines are the 90\% completeness limit assuming
a limit of I$_{\rm E}$ = 20.5.
\label{fig:concmass}}
\end{figure*}

In Figure \ref{fig:concmass}, we plot concentration vs.
log M* for our highly reliable
samples of 2-armed and 3-armed 
galaxies within four redshift ranges:
0.2 $<$ z $\le$ 0.4, 0.4 $<$ z $\le$ 0.6, 0.6 $<$ z $\le$ 0.8,
and 0.8 $<$ z $\le$ 1.  
There are 2361, 2064, 1414, and 485 2-armed galaxies 
and 123, 79, 48, and 12 3-armed galaxies,
respectively, 
in the four redshift bins.
The contours give relative numbers of galaxies per area on the plot,
for
2-armed 
galaxies
(cyan contours) and 3-armed galaxies (black contours).
In Figure \ref{fig:concmass}, 
we have placed vertical
lines marking the 90\% completeness
limit 
using the strict cut-off of I$_E$ = 20.5,
plus the 50\% completeness limit 
with the more relaxed cut-off of I$_E$ = 21.

In Figure \ref{fig:concmass}, we have color-coded the datapoints based on
S\'ersic index.
In all four panels,
there is a trend of S\'ersic
index with concentration: galaxies with larger concentrations
have higher mean S\'ersic indices.
Only a small subset of the galaxies in our sample have S\'ersic
index greater than 3; most of these are 2-armed galaxies.
Very few 3-armed galaxies have S\'ersic indices greater than 2.
The high mass, high concentration galaxies tend
to have larger S\'ersic indices (up to $\sim$4, typical 
of elliptical-like galaxies), while
low concentration galaxies tend low S\'ersic indices,
around 1, as expected from flat disk galaxies.
Even in the high redshift ranges
we see some S\'ersic index = 4 galaxies among the 2-armed
galaxies.

\begin{figure*}[htp]
\includegraphics[width=9.05cm,trim={0.0cm 0cm 6.5cm 15cm},clip=true]{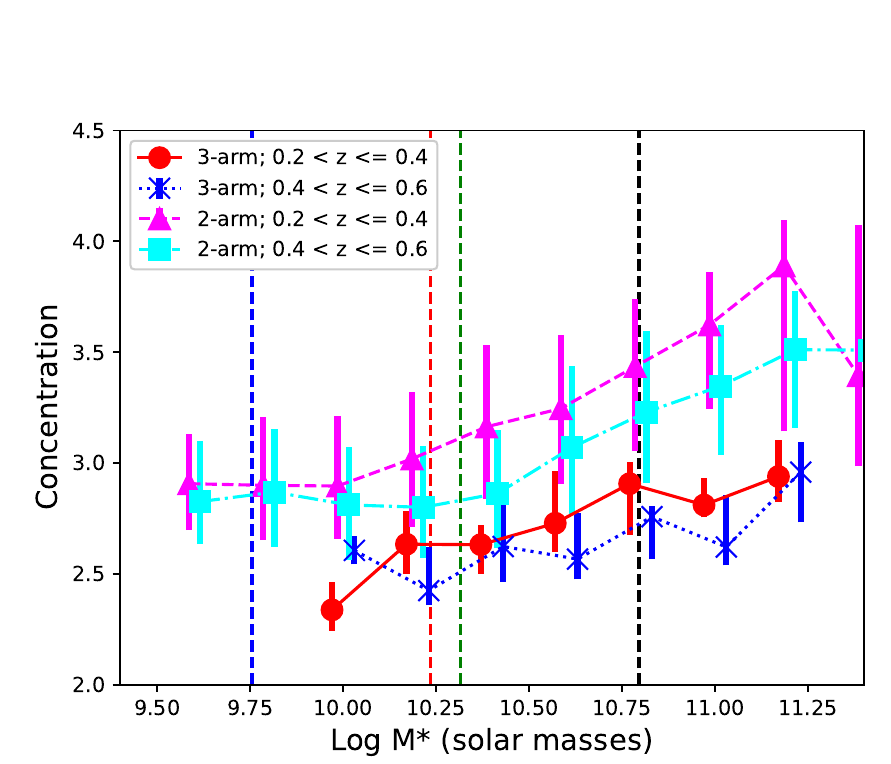}
\includegraphics[width=10.05cm,trim={0.0cm 0cm 4cm 15cm},clip=true]{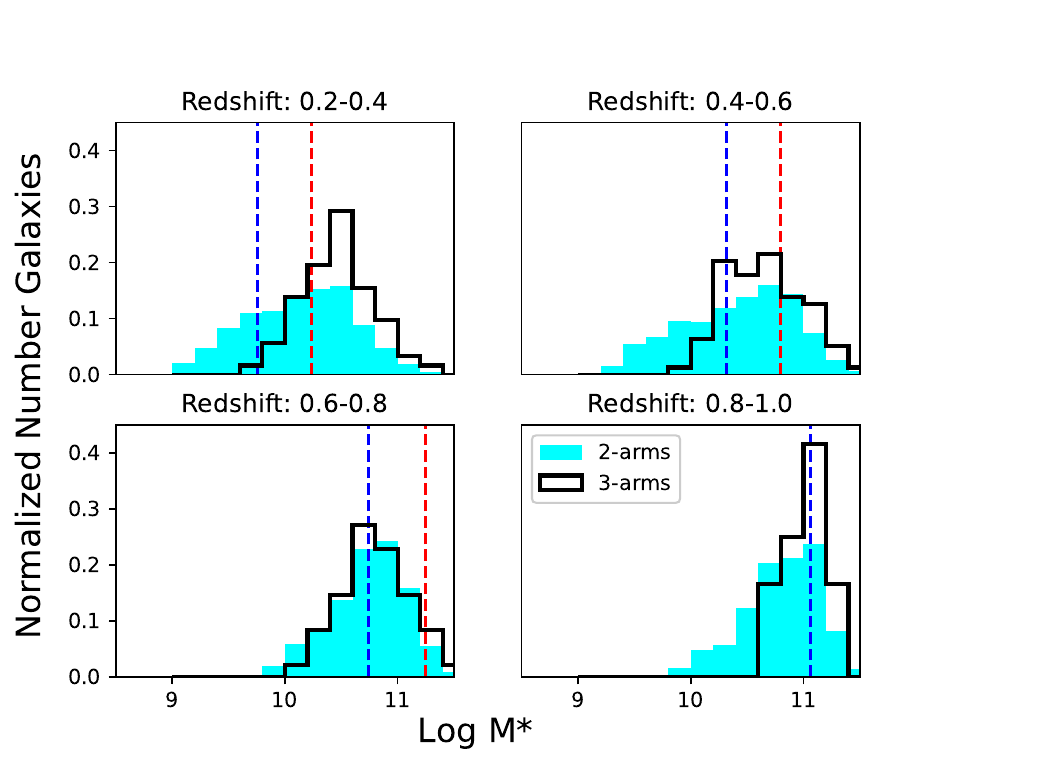}
\caption{
Left: After binning 
into stellar mass bins with bin size $\Delta$(log M*) = 0.2,
this plot displays the 
median concentrations as a function of stellar mass
for 2-armed galaxies in the
0.2 $<$ z $\le$ 0.4 range (cyan squares), 2-armed
galaxies in the 0.4 $<$ z $\le$ 0.6 range (magenta triangles),
3-armed galaxies in the 0.2 $<$ z $\le$ 0.4 range (red circles),
and 
3-armed galaxies in the 0.4 $<$ z $\le$ 0.6 range (blue crosses).
The errorbars display the range from the first to the third quartile.
The datapoints have been slightly offset in the x direction to avoid
overlap.
The dark blue dashed vertical line represents the
50\% completeness limit 
at
0.2 $<$ z $\le$ 0.4 
assuming a magnitude limit of 
I$_{\rm E}$ = 21,
while the red dashed line is the 90\% completeness limit in the
0.2 $<$ z $\le$ 0.4 range
assuming
a limit of I$_{\rm E}$ = 20.5.
The cyan (light blue) dashed vertical line represents the
50\% completeness limit at 0.4 $<$ z $\le$ 0.6 assuming a magnitude limit of 
I$_{\rm E}$ = 21,
while the black dashed line is the 90\% completeness limit at
0.4 $<$ z $\le$ 0.6
assuming
a limit of I$_{\rm E}$ = 20.5.
Right: Histograms of the stellar masses for the 2-armed galaxies
(filled cyan histograms) and the 3-armed galaxies (histograms
outlined in black) in the four redshift ranges.  
Each histogram has been normalized so
that the total area under the curve is equal to 1.  
The dark blue dashed vertical lines represent the
50\% completeness limit 
in the plotted redshift range
assuming a magnitude limit of 
I$_{\rm E}$ = 21,
while the red dashed lines are the 90\% completeness limit in the
redshift range
assuming
a limit of I$_{\rm E}$ = 20.5.
\label{fig:median}}
\end{figure*}

In all four panels in Figure \ref{fig:concmass}, 
for a given
stellar mass the 3-armed galaxies tend to have smaller 
concentrations than the 2-armed galaxies.
We show this same result in a different way
in the left panel of Figure \ref{fig:median}, where
we bin the galaxies into mass bins with bin size $\Delta$(log M*) = 0.2,
and compare the median concentrations for the 2-armed and 3-armed galaxies 
as a function of redshift
for the first and second redshift ranges.
For a given redshift and mass range, 2-armed galaxies tend to have larger concentrations than
3-armed galaxies. 
If Figure
\ref{fig:concmass} and 
Figure \ref{fig:median} were made using a 2-armed vote threshold of $\ge$0.5 rather than $\ge$0.8,
this conclusion would not change.

We ran
Kolmogorov-Smirnov (KS) and Anderson-Darling (AD) tests to search for
significant
differences in the concentrations
of the galaxies.
For the 0.2 $<$ z $\le$ 0.4 redshift range,
we found 
significantly larger concentrations for the 2-armed galaxies
than for the 3-armed galaxies in all the mass bins between
9.9 $\le$ log (M*/M$_{\sun}$) $<$ 10.1 and 10.9 $\le$ log (M*/M$_{\sun}$) $<$ 11.1 
(KS/AD probabilities
of the samples being drawn from the same parent sample are $\le$0.002).
Similar statistics are found when the selection criteria for
2-armed galaxies is relaxed to vote fraction $\ge$ 0.5 rather
than $\ge$0.8.
For the 0.4 $<$ z $\le$ 0.6 range, 
for all the mass bins between 10.1 $\le$ log (M*/M$_{\sun}$) $<$ 10.3 and 
11.1 $\le$ log (M*/M$_{\sun}$) $<$ 11.3 the KS/AD probabilities are $\le$0.006,
with the 2-armed galaxies having larger concentrations.
Similar tests for 
0.6 $<$ z $\le$ 0.8 give significant differences 
between 2-armed and 3-armed galaxies in all mass bins
between
10.5 $\le$ log (M*/M$_{\sun}$) $<$ 10.7 and 
11.3 $\le$ log (M*/M$_{\sun}$) $<$ 11.5, with KS/AD probabilities $\le$0.007.
For the highest redshift range, only the 
11.1 $\le$ log (M*/M$_{\sun}$) $<$ 11.3 mass bin shows significant
difference, with poorer but still significant KS/AD probabilities of 0.02.

The catalogued
concentrations of the 2-armed galaxies in the 0.2 $<$ z $\le$ 0.4 range
are larger than those of 2-armed galaxies
in the 0.4 $<$ z $\le$ 0.6 range
(left panel, Figure \ref{fig:median}).
In the mass bins between 9.9 $<$ log (M*/M$_{\sun}$) $\le$ 10.1
and 11.1 $<$ log (M*/M$_{\sun}$) $\le$ 11.3, 
this difference is significant, with 
KS/AD probabilities $\le$ 0.003,
except in the highest and lowest mass bins, 
which have KS/AD probabilities $\le$ 0.02.
This apparent different in concentration with z
may be an artifact caused by redshift biases.
Two different factors may contribute: 
poorer effective spatial resolution at higher redshifts,
and shorter rest wavelengths at higher redshifts (i.e.,
a morphological
K-correction).
Numerous studies have shown that, if images of nearby
spiral 
galaxies are smoothed to mimic the effect of higher redshifts,
observed concentrations tend to decrease \citep{2021ApJ...919..139W,
2021MNRAS.507..886T,
2023A&A...676A..74Y,
2024A&A...686A.100W,
2025arXiv251109644S}.
Smoothing tends to increase 
r$_{20}$, therefore decreasing the measured
concentration.   
The redshift bias in concentration
has been quantified in these earlier studies in terms of
the galaxy size (either effective radius or r$_{90}$)
divided by either the 
FWHM of the 
point spread function, or the pixel size.
For our sample galaxies, the median Kron radius 
(roughly r$_{90}$; \citealp{2005PASA...22..118G})
is about 6$''$ $-$ 7$''$, and the median
half-light radius is about 1\farcs1 $-$ 1\farcs3,
with little variation between our two redshift bins.
Given the pixel size and resolution of the Euclid images,
published correction tables 
\citep{2024A&A...686A.100W,
2025arXiv251109644S}
imply that the corrections 
to the measured concentrations due to redshift are 
small, $\le$ 0.1, and are not expected to vary too much
with redshift in our redshift range.
For 2-armed galaxies, we see an observed difference in concentration
of about 0.2 between z $\sim$ 0.3 and z $\sim$ 0.5.
This is larger than the expected effect due to resolution,
indicating a possible residual difference in concentration.
To test this in detail, for each 2-armed galaxy in the
sample we applied the 
\citet{2025arXiv251109644S} 
correction prescription to the concentration, and re-ran the
KS/AD tests.  We again obtained low KS/AD probabilities,
indicating significant differences even after this correction.

This calculation, however, does not take into account possible
changes in morphology
with wavelength, which may also contribute to the
observed redshift variations in concentration.  At low redshifts,
we observe longer rest wavelengths than at higher redshifts,
and bulge stars contribute more at longer wavelengths.
The change in concentration with redshift due to this effect
has been quantified by \citet{2007ApJ...659..162T},
who measured concentrations of 199 nearby galaxies in several
UV and optical filters.  The Euclid I$_{\rm E}$ filter
at z = 0.3 has a rest wavelength roughly equivalent to 
the optical V filter, while the rest wavelength
observed by the Euclid filter at z = 0.5
approximately corresponds to the optical B filter.
The median concentrations in the V and B filter
for the Sa $-$ Sc galaxies 
in the \citet{2007ApJ...659..162T} sample are 2.87 and 2.71,
respectively,
giving a differential concentration of 0.13.   
When this offset of 0.13 plus the resolution correction
is added to the observed Euclid concentrations
of the galaxies in the 0.4 $<$ z $\le$ 0.6 redshift range,
the concentrations agree well with galaxies at 0.2 $<$ z $\le$ 0.4
in the same mass range, and KS/AD probabilities show that
we cannot rule out that
the two sets of galaxies come from the same parent
sample.   The only mass bin for which we still see a significant
difference in the concentrations is the
10.3 $\le$ (M*/M$_{\sun}$) $<$ 10.5 bin.  After both corrections were
applied, the galaxies in the lower redshift range still have
significantly higher concentrations (KS/AD probabilities
$<$ 0.001).   
In Section \ref{sec:bulge_growth}.
we discuss these differences further.
In contrast to the 2-armed galaxies, 
KS/AD tests do not show significant differences 
in the concentrations of
the 3-armed galaxies
between these two redshift
ranges. 

In the two lower redshift ranges,
the distribution of stellar masses for the 3-armed galaxies
is shifted to higher masses compared to 
the 2-armed galaxies 
(Figure \ref{fig:concmass}).
In the 0.2 $<$ z $\le$ 0.4
redshift range, 3-armed galaxies tend to be both more
massive and less concentrated than 2-armed galaxies.
At 0.2 $\le$ z $<$ 0.4,
the maxima for both samples are above the 
nominal 50\% mass completeness
limit,  
thus these offsets in mass are probably not caused by
incompleteness.
These results are shown more clearly in the right panel of 
Figure 
\ref{fig:median}, where we provide histograms of the
stellar masses of 2-armed vs.\ 3-armed galaxies for our four
redshift bins.
In the two lowest redshift
ranges, the distribution of stellar masses for 2-armed galaxies
is skewed to lower masses than that of 3-armed galaxies.
KS/AD tests show that these differences are significant
(KS/AD probabilities $\le$ 0.001).
In these two redshift ranges, 
there is a shortage of 3-armed
galaxies below about log (M*/M$_{\sun}$) = 10, relative to 2-armed galaxies.  
This deficiency of low mass 3-armed
galaxies
does not appear to be caused by incompleteness,
given that the completeness limits for 2-armed and 3-armed galaxies 
are similar.
These trends do not change when the 2-arm vote fraction
threshold
is decreased to 0.5.
We discuss possible explanations for the mass difference 
between 2-armed and 3-armed galaxies
in Section
\ref{sec:bend}.
In the 0.6 $<$ z $\le$ 0.8 and 0.8 $<$ z $\le$ 1.0 redshift
ranges, KS/AD tests do not show significant differences in 
the mass distributions of the two samples, perhaps because
of incompleteness at the low mass end.

In the first two panels of Figure \ref{fig:concmass}, there is a clear bend in
the concentration-vs-stellar mass relation for 2-armed
galaxies.  Above about log (M*/M$_{\sun}$) = 10.3, a population
of high concentration 2-armed galaxies exists.  
For 3-armed
galaxies, higher mass galaxies 
tend
to have larger concentrations, 
however, the relation appears linear; no bend is seen. 
Above log (M*/M$_{\sun}$) = 10.3, the low concentration galaxies in our sample are mostly
3-armed rather than 2-armed.
In the 0.6 $<$ z $\le$ 0.8 and 0.8 $<$ z $\le$ 1.0 samples of 2-armed
galaxies, 
there is still an apparent bend in the relation for 2-armed
galaxies. 

\subsection{The Star-Forming Galaxy Main Sequence} \label{sec:main_seq}

\begin{figure*}[htp]
\includegraphics[width=16.0cm,trim={0.0cm 0cm 6cm 15cm},clip=true]{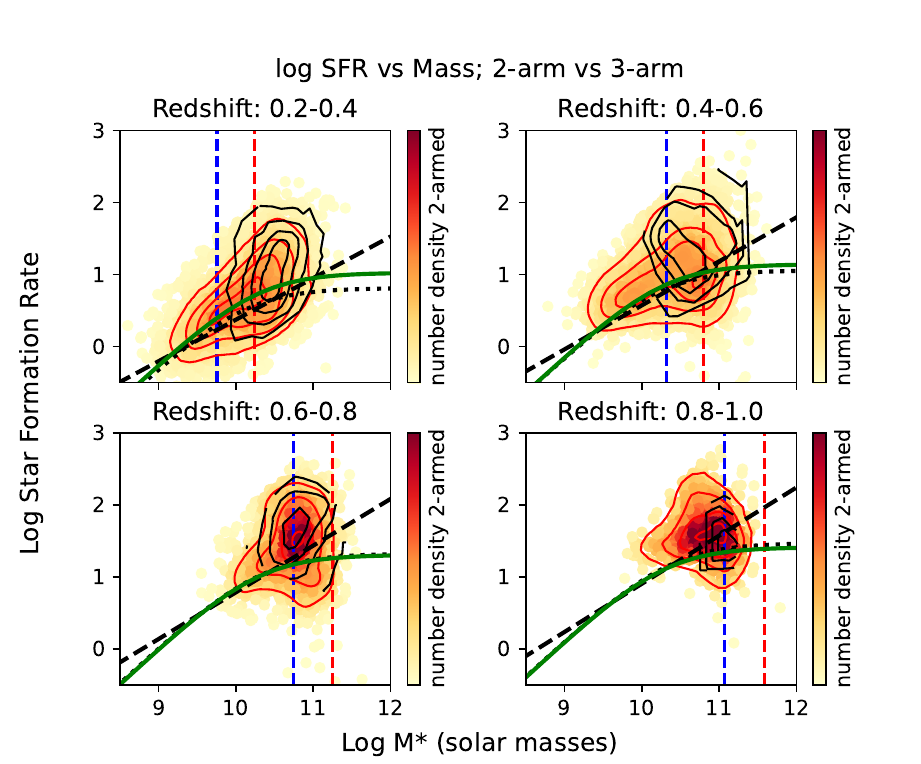}
\caption{SFR (in solar masses per year) vs. log M* for four redshift ranges,
for galaxies in our highly reliable samples
of 2-armed and 3-armed galaxies.
The contours give the relative number density of 2-armed
(red contours) and 3-armed galaxies (black contours) per area on the plot.
The background color scale gives the relative number density of 2-armed
galaxies on the plot, as shown in the color scales on the right.
The blue dashed vertical lines represent the
50\% completeness limit assuming a magnitude limit of 
I$_{\rm E}$ = 21,
while the red dashed vertical line is the 90\% completeness limit assuming
a limit of I$_{\rm E}$ = 20.5.
The star-formation main sequence from 
\citet{2014ApJS..214...15S}
is
plotted as a black dashed line, while that of 
\citet{2023MNRAS.519.1526P}
is the dotted
black
curve.
The solid green curve is the main sequence from
\citet{2025arXiv250315314E}.
\label{fig:SFRmass_concentration}}
\end{figure*}

The galaxy main sequence
is a well-known correlation between log SFR
and log M* for star-forming galaxies
\citep{2007ApJ...660L..43N, 
2007ApJS..173..267S}. 
In Figure \ref{fig:SFRmass_concentration}, 
we plot log SFR vs.\ log M* for our 
highly reliable 2-armed
and 3-armed galaxies, dividing
the galaxies into our four redshift ranges. 
We mark completeness limits
by vertical lines.
In making Figure \ref{fig:SFRmass_concentration}, 
we excluded galaxies with
SFR $<$ 10$^{-2.5}$ M$_{\sun}$~yr$^{-1}$ (i.e., 
log sSFR $<$ $-$12 for M* $\ge$ 10$^{9.5}$~M$_{\sun}$), since
sSFR at these low levels are very uncertain
\citep{2007ApJS..173..315S, 2016ApJS..227....2S}.
Galaxies with log SFR $<$ $-$2.5 constitute $<$3$\%$ of the galaxies
in the 0.2 $<$ z $\le$ 0.4 range.
When 
Figure \ref{fig:SFRmass_concentration} is created using
a 2-arm vote fraction threshold of 0.5 rather than 0.8, its appearance
changes only slightly.

In Figure \ref{fig:SFRmass_concentration}, 
in the 0.2 $<$ z $\le$ 0.4 range
the distribution of SFRs peaks at higher SFRs for the 3-armed galaxies.
However, the 3-armed galaxies also have higher stellar masses.
In the lowest redshift range,
the 2-armed
and 3-armed galaxies make a single continuous sequence,
with the distribution of 3-armed galaxies
peaking at higher SFRs and higher stellar masses
than the 2-armed galaxies.  
With increasing redshift, the bottom of the main sequence 
disappears because of incompleteness.

To compare the SFRs of the 2-armed and 3-armed galaxies more quantitatively,
we binned the galaxies into $\Delta$(log M*) = 0.2 mass bins
and ran KS/AD tests comparing the SFRs of the 2-armed and 
3-armed galaxies at a given redshift and in a given mass bin.
When comparing galaxies with similar stellar
masses and redshifts, 2-armed and 3-armed galaxies have similar
SFRs, except in a few mass ranges in the higher redshift
bins, where the 3-armed galaxies have higher SFRs on average.
In the 0.2 $<$ z $\le$ 0.4 range,
we see no significant difference
between 2-armed and 3-armed galaxies in any mass range.
In the 0.4 $<$ z $\le$ 0.6 range, 
in two mass bins
(10.3 $\le$ log (M*/M$_{\sun}$) $<$ 10.5 and 
10.7 $\le$ log (M*/M$_{\sun}$) $<$ 10.9), 
the 3-armed galaxies have significantly
larger SFRs than the 2-armed galaxies
(KS/AD probabilities
$\le$0.001 and $\le$0.03 for the two bins,
respectively).
In the 0.6 $<$ z $\le$ 0.8 range, 
in only one mass bin 
(11.1 $\le$ log (M*/M$_{\sun}$) 
$<$ 11.3)
is there a marginally significant difference
(KS/AD probabilities of 0.04/0.05),
again with the 3-armed galaxies having higher
SFR.  No clear differences are seen in the
highest redshift range.

In Figure \ref{fig:SFRmass_concentration}, 
we overlay published main sequences 
\citep{2014ApJS..214...15S, 
2023MNRAS.519.1526P,
2025arXiv250315314E}
on the Euclid plots. 
In the lowest redshift range,
the relation
for our spirals 
agrees well with the earlier Euclid result
\citep{2025arXiv250315314E}, but 
lies slightly above other published main sequence
relations  
\citep{2014ApJS..214...15S, 
2023MNRAS.519.1526P},
with SFRs enhanced by a factor of $\sim$2 compared to their main
sequence.
Most of the 2-armed and 3-armed galaxies in our sample are close
to
the main sequence or above it, but in the middle
redshift ranges, a few high mass 2-armed galaxies lie below the published
main sequence.
Although some
red spirals have been identified, they are relatively
rare 
\citep{2010MNRAS.405..783M}.
Once star formation
has dimmed, we cannot easily pick out spiral patterns.   

\begin{figure*}[htp]
\includegraphics[width=16.0cm,trim={0.0cm 0cm 6cm 15cm},clip=true]{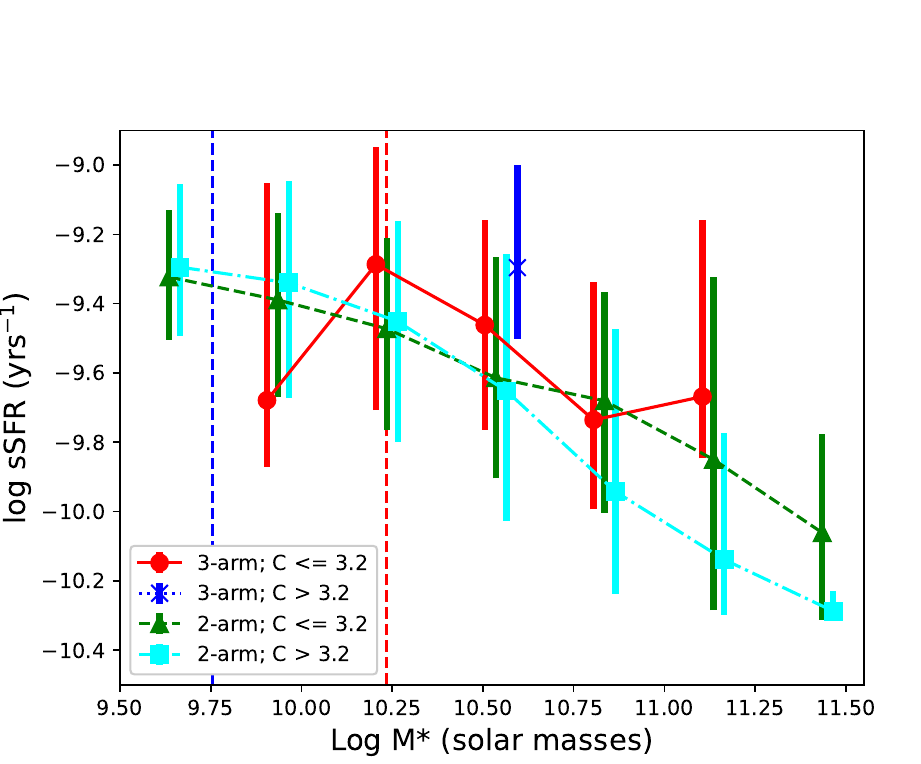}
\caption{
After binning 
into stellar mass bins with bin size $\Delta$(log M*) = 0.3 and dividing
into two concentration bins,
this plot displays the 
median sSFR as a function of stellar mass
for 2-armed galaxies vs.\ 3-armed galaxies
in the
0.2 $<$ z $\le$ 0.4 range.
The errorbars display the range from the first to the third quartile.
The datapoints have been slightly offset in the x direction to avoid
overlap.
The dark blue dashed vertical lines represent the
50\% completeness limit 
at
0.2 $<$ z $\le$ 0.4 
assuming a magnitude limit of 
I$_{\rm E}$ = 21,
while the red dashed lines are the 90\% completeness limit in the
0.2 $<$ z $\le$ 0.4 range
assuming
a limit of I$_{\rm E}$ = 20.5.
This plot excludes datapoints 
for bins with two or less galaxies.
\label{fig:SFRmass_concentration_2nd}}
\end{figure*}

To test whether SFR is a function of concentration
at fixed stellar mass for the galaxies in our sample,
we divided our 2-armed and 3-armed galaxies into two 
groups by concentration (C $\le$ 3.2 and C $>$ 3.2),
and then binned the galaxies by mass into bins with
width log M* = 0.3.  
Figure \ref{fig:SFRmass_concentration_2nd}
plots the median concentration in each subset as a function
of log M*.  
We then ran KS/AD tests in each M* bin, 
comparing the sSFRs for 
high and low concentration galaxies for a given
arm count.
Since there 
are few high concentration 3-armed galaxies
in our sample, for most mass bins we are only
comparing 2-armed vs.\ 2-armed galaxies in
the two concentration ranges.
For only one bin did we find a significant difference:
in the 10.7 $\le$ log (M*/M$_{\sun}$) $<$ 11.0 bin,
the sSFRs for high concentration 2-armed galaxies
are significantly lower than those of low concentration
galaxies (KS/AD probabilities of 0.0013/0.0010, respectively).
This suggests that at the high mass end of the main sequence,
there may be some quenching of 2-armed spirals with bulge growth.

\section{One-Armed Galaxies} \label{sec:1arm}

\begin{figure*}[htp]
\includegraphics[width=16.0cm,trim={0.0cm 0cm 6cm 15cm},clip=true]{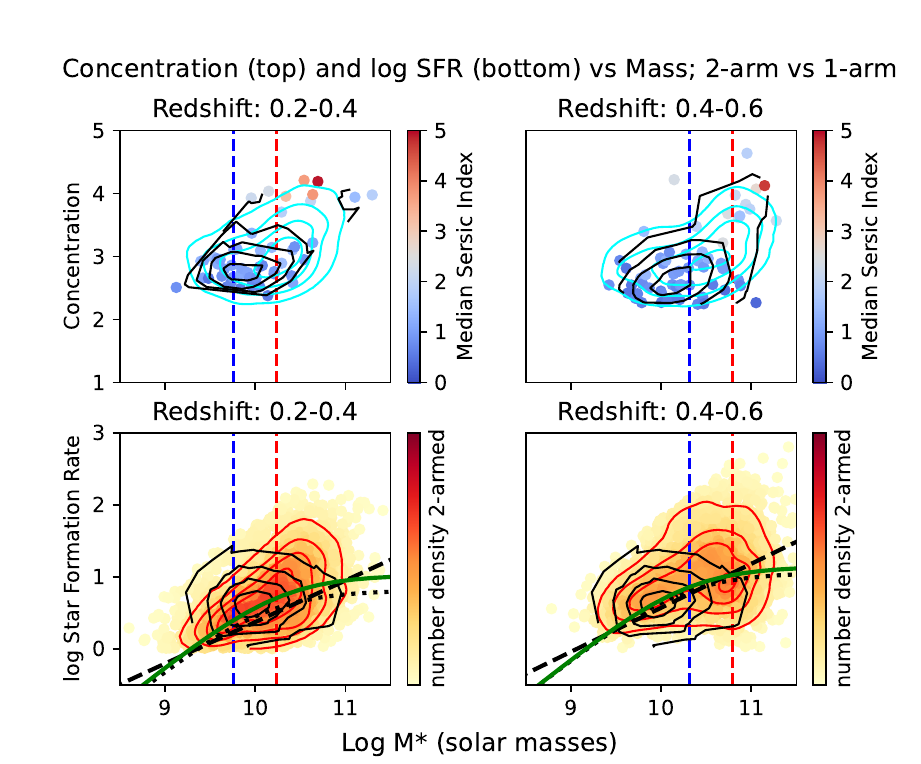}
\caption{Top row: Concentration vs.\  log M* for 0.2 $<$ z $\le$ 0.4 (left)
and 0.4 $<$ z $\le$ 0.6 (right),
for galaxies in our highly reliable samples
of 2-armed and 1-armed galaxies.
The contours give the relative number of galaxies per area on the plot
for
2-armed (cyan contours) and 1-armed galaxies (black contours).
The background color scale gives the S\'ersic index of the 1-armed
galaxies on the plot, as shown in the color scales on the right.
Bottom row: Log SFR (in solar masses per year)
vs.\  log M* for 0.2 $<$ z $\le$ 0.4 (left)
and 0.4 $<$ z $\le$ 0.6 (right).
The contours give the relative number of galaxies per area on the plot
for
2-armed (red contours) and 1-armed galaxies (black contours).
The color scale displays
the relative number density of 2-armed galaxies on the plot.
In all four panels,
the blue dashed vertical lines represent the
50\% completeness limit assuming a magnitude limit of 
I$_{\rm E}$ = 21,
while the red dashed vertical lines are the 90\% completeness limit assuming
a limit of I$_{\rm E}$ = 20.5.
In panels in the bottom row,
the star-formation main sequence from 
\citet{2014ApJS..214...15S}
is
plotted as a black dashed line, while that of 
\citet{2023MNRAS.519.1526P}
is the dotted
black
curve.
The solid green curve is the main sequence from
\citet{2025arXiv250315314E}.
\label{fig:onearmed}}
\end{figure*}

In Figure \ref{fig:onearmed}, we compare our
highly reliable samples of 1-armed and 2-armed galaxies.
In the top row of 
Figure \ref{fig:onearmed}, we plot concentration
vs.\ log M* for 
0.2 $<$ z $\le$ 0.4 (left panel)
and 0.4 $<$ z $\le$ 0.6 (right panel).
These plots show that 1-armed galaxies
generally have lower stellar masses than 2-armed galaxies,
however there is considerable scatter.  
For a given stellar mass, the concentrations
and SFRs of 1-armed galaxies are similar to those of
2-armed galaxies.
After binning the data into $\Delta$(log M*) = 0.2 mass bins,
KS/AD tests show no significant differences in the concentrations
of 
1-armed and 
2-armed galaxies with the same mass,
except in 
the
10.5 $\le$ log (M*/M$_{\sun}$) $<$ 10.7 mass range at
0.2 $<$ z $\le$ 0.4, 
where the one armed galaxies have 
larger concentrations, a result that is 
marginally significant (KS/AD probabilities of 0.04/0.03,
respectively).
Also, in the 9.7 $\le$ log (M*/M$_{\sun}$) $<$ 9.9 mass bin for 
0.4 $<$ z $\le$ 0.6, 2-armed galaxies have marginally larger
concentrations (KS/AD probabilities of 0.01 and 0.02, respectively).

In the top panels
of Figure \ref{fig:onearmed}, 
the datapoints shown are the values for the 1-armed galaxies
after color-coding to the S\'ersic index.   The 1-armed 
galaxies tend to have low S\'ersic indices, $\sim$ 1,
typical of flattened disks, however, a handful of
1-armed galaxies with large concentrations also have
large S\'ersic indices.  

In the bottom row
of Figure \ref{fig:onearmed}, 
we plot log SFR vs.\ log M* for 
1-armed vs.\ 2-armed galaxies at
0.2 $<$ z $\le$ 0.4 (left panel)
and 0.4 $<$ z $\le$ 0.6 (right panel).
In the lower redshift range, KS/AD tests show no significant
difference between the SFRs of 2-armed and 1-armed galaxies in the
same mass range.  
In the 0.4 $<$ z $\le$ 0.6 range, the 1-armed galaxies have significantly
larger SFRs than 2-armed galaxies of the same mass only for
the  
10.3 $\le$ (log M*/M$_{\sun}$) $<$ 10.5 and 
10.7 $\le$ log (M*/M$_{\sun}$) $<$ 10.9 
mass bins (KS/AD probabilities $\le$0.001 and $\le$0.003, respectively).

\section{Discussion} \label{sec:discussion}

\subsection{Arm Counts vs.\ Concentration, M*, and SFR} 

Using data from the Euclid catalogs, we find 
2-armed galaxies have larger concentrations than 3-armed galaxies
for a given stellar mass.   
This result has been seen before in 
more nearby galaxies 
\citep{1982MNRAS.201.1021E, 2017MNRAS.471.1070B,
2020ApJ...900..150Y,
2024AJ....168...12S}.
With Euclid, 
we have extended this trend out to
a redshift of 1.
We also find that 2-armed galaxies tend to have larger S\'ersic indices
than 3-armed galaxies.
This result is not surprising since S\'ersic
index is well-correlated with concentration 
\citep{2025arXiv250315309E}.
This result is consistent with expectations of spiral density wave
theory, which
suggests that a large bulge may
stabilize
a spiral density wave, allowing a 
2-armed spiral pattern to persist 
\citep{1985IAUS..106..513L, 1989ApJ...338...78B, 1989ApJ...338..104B, 2016ApJ...826L..21S}.
Two-armed spirals are also created in 
simulations of disk instabilities in isolated galaxies,
if the disk mass is
large compared to the halo and the bulge is large enough
\citep{2013A&A...553A..77G, 2015ApJ...808L...8D, 2018MNRAS.481..185M, 2025Galax..13..132P}.
Galaxies with large bulges tend to have declining rotation curves, which in 
turn produces a large shear rate,
which favors 2-armed morphologies
\citep{2013A&A...553A..77G, 2015ApJ...808L...8D, 2018MNRAS.481..185M, 2025Galax..13..132P}.
Tidal interactions with low mass companions can also generate 
2-armed patterns \citep{1992AJ....103.1089B},
even in very low density environments
\citep{2025arXiv250722793Q}.
The presence of a bulge 
may allow a 2-armed 
pattern to last longer, whether the spiral structure is triggered by an interaction
or by a random fluctuation in the disk.
\citet{2015ApJS..217...32B}
distinguish between normal grand design galaxies
and a subset of systems they call `extreme spirals'.   They define
`extreme spirals'
as 2-armed galaxies with large amplitude arms and
very open spirals.  They suggest that many
of these could be tidal spirals.    This concept of
two distinctly different types of 2-armed spirals deserves
more exploration in future studies.

In the lowest redshift range studied in this survey, 0.2 $\le$ z $<$ 0.4, 
we see a difference in the stellar masses of 2-armed and 3-armed galaxies,
with 3-armed galaxies being more massive on average.  
This is consistent with SDSS 
studies of more nearby galaxies 
\citep{2024AJ....168...12S}.
Why such a mass dependence exists is discussed below in
Section 
\ref{sec:bend}.
Because of sample incompleteness 
we cannot make clear 
conclusions about mass differences between 2-armed and 3-armed
galaxies at z $>$ 0.4.

At 0.2 $<$ z $\le$ 0.4, 
the 3-armed galaxies tend to have higher SFRs
than the 2-armed galaxies, but
for a given stellar mass 2-armed and 3-armed galaxies
have similar SFRs.  
This result is not surprising given the
well-known main sequence
correlation between SFR and stellar mass for star-forming galaxies
\citep{2007ApJ...660L..43N, 
2007ApJS..173..267S}
and the tendency of 3-armed galaxies to have larger stellar masses
than 2-armed galaxies.
This result is consistent with earlier studies at lower redshifts.
Using Galaxy Zoo 2 arm counts for 
SDSS galaxies, 
\citet{2017MNRAS.468.1850H} and \citet{2017MNRAS.472.2263H}
found similar sSFRs for 2-armed and 3-armed galaxies, for stellar-mass-matched
samples.
\citet{2017MNRAS.468.1850H} conclude that their results support
a scenario in which spiral arms mark the distribution
of star-forming gas
in galaxies, but are not directly responsible for
star formation triggering (e.g., \citealp{2002ApJ...577..206E}).
\citet{2015MNRAS.449..820W} also found that the slope and dispersion
of the galaxy main
sequence for SDSS galaxies is independent of arm count.
In addition, \citet{1986ApJ...311..554E} found no significant variations
in the SFR per area in the disks of spiral galaxies as a function
of arm class.
\citet{2024AJ....168...12S}
found no significant
difference between the sSFRs of grand design and multi-armed galaxies,
when comparing galaxies with similar stellar masses and concentrations.
However, grand design galaxies tend to have larger concentrations,
and galaxies with larger concentrations tend to have lower sSFR.
Similarly, 
\citet{2022AJ....164..146S}
found a weak correlation between the m=3 Fourier amplitude of
spiral galaxies and the sSFR, but this disappeared when comparing
galaxies with similar M* and concentration.

In the higher redshift ranges,
in a few mass ranges near the top of the main sequence
the 3-armed galaxies have enhanced SFRs
compared to 2-armed galaxies of the same stellar mass.  
This result needs to be confirmed with larger more complete samples
and more reliable photometric redshifts.
If confirmed, one 
possible explanation might be 
that gas inflow into the galactic center 
has triggered a central starburst in the 3-armed galaxies,
which may eventually lead to a more massive bulge.
This supports the idea that 3-armed morphologies
are favored in turbulent gas-rich galaxies
\citep{2024ApJ...968...86B,
2025A&A...700A..42E},
and suggests that disk gas fraction may be a factor
in the evolution of spiral patterns in galaxies.
Followup work with larger samples is needed to fully explore these issues.

\citet{2025arXiv250315309E}
investigated the
S\'ersic indices of
1,312,068
Euclid Q1 galaxies with I$_{\rm E}$ $<$ 23.
They identify a large population of S\'ersic index $\sim$ 4 early-type
galaxies
below the main sequence between 10 $\le$ log (M*/M$_{\sun}$) $\le$ 11. 
Along the main sequence, they find mostly S\'ersic index $\sim$ 1 
up to about log (M*/M$_{\sun}$) = 10.5, above which
the S\'ersic index increases to a median
of $\sim$ 2 by log (M*/M$_{\sun}$) = 11.  In this regime (high mass main sequence
galaxies), 
\citet{2025arXiv250315309E}
detect a large dispersion in the S\'ersic index of
the galaxies, suggesting that morphologies change
while the galaxies are still star-forming.
This is consistent with earlier 
statistical studies of galaxy populations over
time which find that star formation quenching is
correlated with bulge growth 
\citep{2010ApJ...709.1018V, 2015ApJ...803...26P}.
However, whether quenching happens before, after, or
during morphological change may depend upon environment
\citep{2025arXiv251102964E}.
Our data suggest that,
at the beginning of this quenching process,
galaxies may be low concentration
multi-armed galaxies near the top of the main sequence.  
Then, after a merger or other bulge-building process, 
the concentration and S\'ersic index increases, 
and the galaxy goes through a phase
where a 2-armed morphology is favored but the galaxy
is still star-forming. In the final
stage, the galaxy 
is completely quenched and has an early-type morphology.

We also compare the concentrations, stellar masses, and SFRs of 1-armed
galaxies with 2-armed galaxies.   Only about 1\% of the
spirals in our sample are classified as 1-armed.
The 1-armed galaxies tend to have lower stellar masses than 2-armed
galaxies, but with some scatter.   The concentrations and SFRs
of the 1-armed galaxies 
are similar to those of 2-armed galaxies with the same stellar mass.
Simulations suggest that leading 1-armed spiral arms can be
produced by retrograde interactions with a companion of
similar mass
\citep{1989A&A...211...25T, 1993ApJ...417..502H}.
This would tend to bias 1-armed galaxies towards lower stellar
masses than galaxies with two spiral arms, since
grand design patterns can be produced by interactions with
lower mass companions (e.g., \citealp{
1992AJ....103.1089B}).
For an SDSS sample of galaxies, \citet{2013MNRAS.429.1051C}
found that galaxies identified as 1-armed in Galaxy Zoo 2
were more likely to have close companions than a control
sample, supporting the hypothesis that these
are predominantly produced by tidal interactions.
Our observation that 1-armed galaxies have similar concentrations
and SFRs to 2-armed galaxies also supports the idea that m=1 structures
are not a function of internal structural parameters,
but instead are linked to tidal interactions.
Followup study is needed of the 1-armed galaxies in this Euclid sample
to confirm their morphologies, investigate their environments,
and search for possible companions.

\subsection{The Bend in the C vs. log M* Relation} \label{sec:bend}

We see a bend in the concentration-vs-mass relation
for 2-armed galaxies.
In our lowest redshift bin, the bend occurs at about log (M*/M$_{\sun}$) = 10.3.  
In contrast,
for 3-armed systems the concentration-mass relation is approximately 
linear, without a bend.
At log (M*/M$_{\sun}$) $>$ 10.5, we see a dichotomy among our spirals:
low concentration galaxies have 3 arms, while high concentration
galaxies have 2 arms.
Such a
bend has been seen before 
at lower redshifts by 
\citet{2020MNRAS.493.1686L}
for disk galaxies in general,
and for 
grand design galaxies by 
\citet{2024AJ....168...12S}.
\citet{2020MNRAS.493.1686L}
suggest that the bend in the C-log M* relation
of spirals 
is caused by the development of classical bulges
in galaxies with higher overall stellar masses.  
Mergers between galaxies can build up the bulge of a spiral,
potentially creating a classical bulge 
\citep{2010ApJ...715..202H, 
2015MNRAS.451....2S, 2017ApJ...840...79S}
and therefore increasing the concentration. 
Secular processes such
as bar-driven gas inflow followed by central
star formation may also help to grow a bulge
\citep{2004ARA&A..42..603K,
2005MNRAS.358.1477A,
2014RvMP...86....1S}.
As discussed above, simulations suggest that 
a classical bulge is more able to maintain
a grand design spiral pattern 
\citep{2016ApJ...826L..21S}.

Perhaps not coincidentally, the bend in the C-log M* relation
lies close to the stellar mass 
at which
the stellar-to-halo-mass function for galaxies in the local
Universe reaches a maximum 
\citep{2010ApJ...717..379B, 2020A&A...634A.135G}.
Galaxies with stellar masses 
higher or lower than about log (M*/M$_{\sun}$) = 10.5 show larger proportions of 
dark matter relative to stars.
According to simulations, 
above this mass threshold
the dominant method of galaxy growth 
is mergers, while below this threshold galaxy growth 
is dominated by gas accretion and star formation 
\citep{2010ApJ...718.1001B}.
Above this threshold mass, star formation may be inhibited by 
virial shock heating of gas falling into a massive halo 
\citep{2003MNRAS.345..349B, 2005MNRAS.363....2K, 
2006MNRAS.368....2D, 
2006MNRAS.370.1651C}
and/or 
AGN feedback 
\citep{2006MNRAS.365...11C, 2008MNRAS.391..481S}.
Below this mass threshold, 
baryons are selectively evicted from the galaxy by
stellar winds
\citep{2003MNRAS.344.1131D},
producing
a correlation between 
M*/M$_{halo}$
and 
total mass.

\begin{figure*}[htp]
\plotone{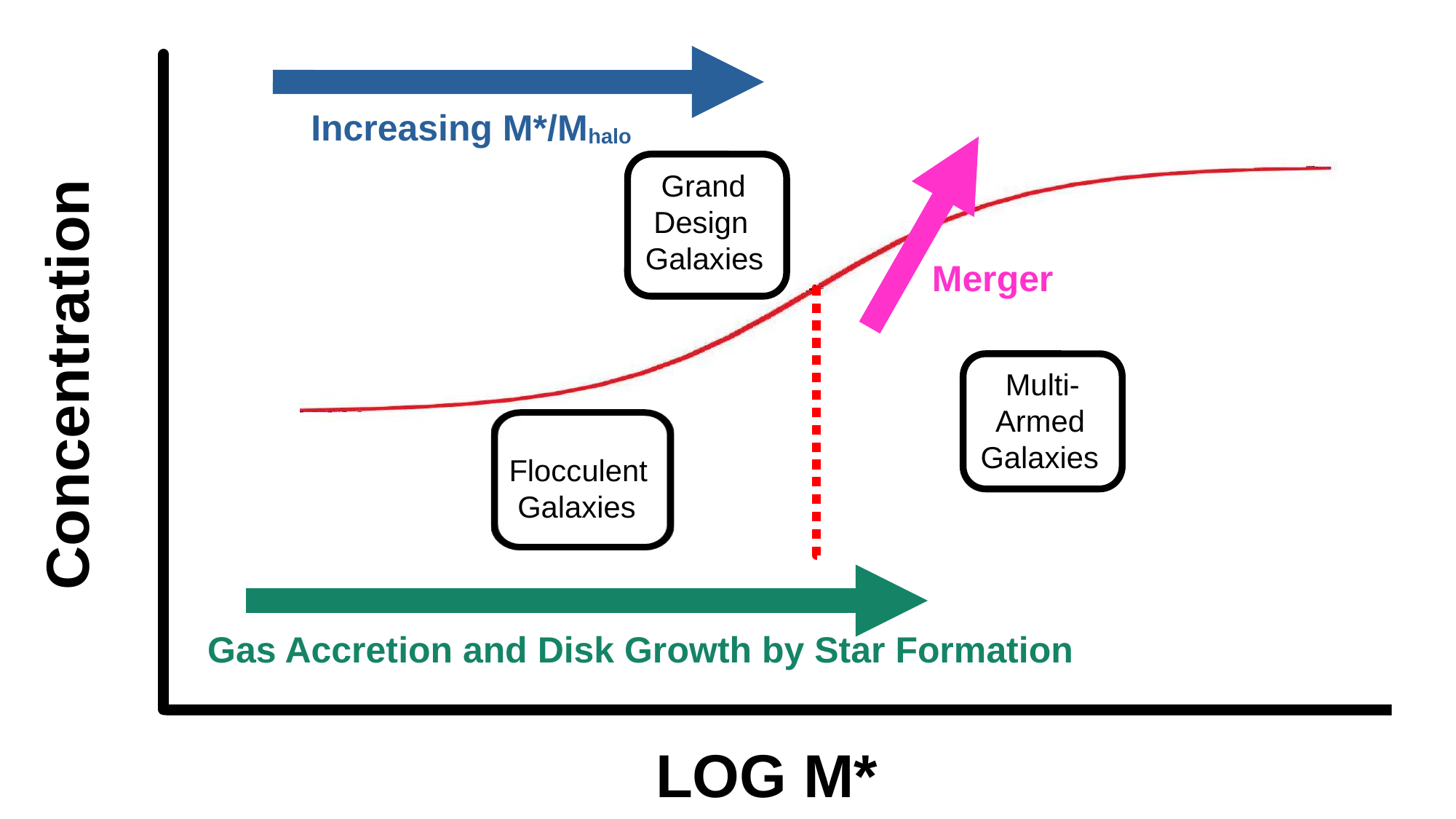}
\caption{
A schematic diagram of the concentration vs.\ log M* plane
for spiral galaxies.
Tentative zones of importance for
2-armed (grand design), multi-armed, and flocculent galaxies
are delineated by the curving red line and the dotted vertical line.
Two-armed galaxies tend to have large concentrations with
a range of masses, but lower concentrations at lower masses.
Multi-armed galaxies tend to have low
concentrations and high stellar masses, while flocculent
galaxies are thought to have low stellar masses and low concentrations.
As illustrated by the blue arrow pointing to the right,
the ratio of stellar mass to halo mass M*/M$_{halo}$ tends to increase
with increasing stellar mass, up to a maximum at about log (M*/M$_{\sun}$)
= 10.5 
\citep{2020A&A...634A.135G}
(dotted red vertical line).
On the left side of the plot, growth in the stellar
mass of a galaxy is thought to be caused primarily by gas
accretion followed by disk star formation, as shown by
the green arrow.  
The magenta diagonal arrow represents the expected direction
of evolution due to a merger, which is expected to
increase the size of the 
bulge and therefore the concentration, and also
increase the stellar mass.
\label{fig:spiral_evolution}}
\end{figure*}

In 
Figure 
\ref{fig:spiral_evolution},
we provide a schematic diagram of concentration vs.\ log M*
for spiral galaxies, and indicate approximately the zones in which 
each of the
three arm classes of galaxies are expected to dominate.
Because of how our samples were selected, flocculent
galaxies are largely omitted from our study.
Earlier studies found that flocculent galaxies tend to be low mass,
compared to grand design and 
multi-armed galaxies 
\citep{2017MNRAS.471.1070B}.
They also tend to
have late Hubble types 
\citep{2011ApJ...737...32E, 2013JKAS...46..141A},
and therefore low concentrations. 
In a plot of C vs.\ log M*, flocculent galaxies
are therefore expected to lie in the lower left corner.

Moving horizontally in Figure
\ref{fig:spiral_evolution}
from the lower left towards larger masses
while keeping the concentration fixed, 
M*/M$_{halo}$ would
tend to increase, according to the stellar-mass-halo-mass 
function
\citep{2020A&A...634A.135G}, and therefore the ratio
of disk mass to halo mass M$_{disk}$/M$_{halo}$ would increase.
Presumably this is the evolutionary path for a low mass, low concentration
galaxy that accretes gas and forms disk stars.
According to numerical simulations,
a larger 
M$_{disk}$/M$_{halo}$
favors multi-armed morphologies
rather than flocculent patterns 
\citep{2013ApJ...766...34D, 2018MNRAS.477.1451F}.
This is consistent with our
observation that 3-armed systems (i.e., multi-armed galaxies)
tend to be massive but have low
concentrations.  
This may be the reason that multi-armed galaxies
tend to have high stellar masses; galaxies with lower stellar
masses will have proportionally larger dark matter haloes, which 
favor flocculent rather than multi-armed morphologies.

Moving vertically upwards from the lower
right in Figure 
\ref{fig:spiral_evolution}
towards a larger
concentration and larger bulge
at a fixed stellar mass and fixed halo mass
would favor 2-armed patterns over
a multi-armed morphology.
This holds for both spirals induced by 
disk instabilities and swing amplification
\citep{2015ApJ...808L...8D, 
2018MNRAS.481..185M}
and long-lived spiral density waves 
\citep{2016ApJ...826L..21S}.
In a merger, both the bulge and 
the overall stellar mass of the galaxy will grow,  
causing the system to travel in an diagonal path upwards and to the
right in Figure 
\ref{fig:spiral_evolution}.
Since bulge mass is correlated with black hole mass 
\citep{2000ApJ...539L...9F, 2003ApJ...596L..27K},
and because 
the probability of radio-loud AGN activity 
increases 
as the mass of a black hole increases
\citep{2006MNRAS.365..101M, 2020ApJ...890..144G},
AGN quenching
is more likely in galaxies with larger bulges. 
Alternatively, a growing bulge may quench star formation
via `morphological quenching' 
\citep{2009ApJ...707..250M},
a process in which a massive
bulge
stabilizes 
a gaseous disk against gravitational collapse,
leading to a lower SFR.
After a galaxy is quenched, the stellar mass only grows via
mergers, thus galaxies tend to move diagonally upwards
in Figure
\ref{fig:spiral_evolution} after quenching.
This scenario of 
gas accretion, star formation, feedback, quenching, and mergers
may be responsible for the bend in the concentration-mass relation
and the $\Sigma$$_1$-mass relation
\citep{2020MNRAS.493.1686L, 2017ApJ...840...47B, 
2022ApJS..262...54G,
2025MNRAS.537.3929C}.

We suggest that, once a disk galaxy has passed the threshold
of log (M*/M$_{\sun}$) = 10.5, it has two possible future evolutionary
paths.  
If it does not
undergo mergers or otherwise fatten its bulge, 
it will retain its low concentration, 
and will tend to have 3 or more arms.  
If it undergoes one or more mergers or grows its bulge by
another process,
then it becomes a 2-armed
spiral, or an elliptical or S0 galaxy.  
If the galaxy exceeds this mass threshold and grows a sufficiently large
bulge, star formation may start to slow down.  We see evidence for this in Figure
\ref{fig:SFRmass_concentration_2nd},
where large-concentration 2-armed spirals 
in the 10.7 $\le$ 
log (M*/M$_{\sun}$) $<$ 11.0 
mass range 
have significantly lower
sSFR than low-concentration 2-armed spirals in the same mass bin.
At lower masses, we do not see differences in sSFR between low and high concentration
2-armed galaxies at fixed mass.   
This supports the idea that bulge growth aids quenching in spirals,
but this quenching only becomes important at high stellar masses, above the bend
in the C vs log M* plot.
At masses below 
the 
log (M*/M$_{\sun}$) = 10.5 threshold,
most spiral galaxies tend to lie on
the same main sequence, 
independent of the number of spiral arms.
This result is consistent with SDSS studies, that showed that most
red spiral galaxies are massive; few are found below our threshold
mass 
\citep{2010MNRAS.405..783M,
2018MNRAS.474.1909F}.
Although low mass galaxies below the main
sequence tend to have larger 
S\'ersic 
indices than main sequence galaxies 
\citep{2011ApJ...742...96W},
implying
larger concentrations, they generally do not have prominent spiral patterns.
In a study of low mass red spirals, \citet{2018MNRAS.474.1909F}
found that they were only located in clusters, suggesting
that they are the products of quenching caused by ram pressure stripping in 
dense environments.   
In less extreme environments, galaxies will
retain their gas and therefore their spiral pattern, and will continue to evolve
up the main sequence, only quenching near the high mass end.

One possible test of these ideas is to determine the 
M*/M$_{halo}$ function for 2-armed and 3-armed galaxies
independently.  
One might expect lower M*/M$_{halo}$ ratios in galaxies with 
larger concentrations if they suffered more mergers over their lifetimes, since mergers add dark matter as well as additional stars.   
Thus 2-armed galaxies might tend to have lower M*/M$_{halo}$ ratios on average than 3-armed galaxies, for the same stellar mass.
Conversely, if stellar mass growth in 3-armed/multi-armed galaxies has
been dominated by gas accretion followed by star formation,
they would be expected to have larger M*/M$_{halo}$ ratios.

At low masses, whether a galaxy is 2-armed or flocculent
may depend upon the central mass of the galaxy.
If a galaxy starts in the lower left of Figure 
\ref{fig:spiral_evolution}
as a flocculent disk
galaxy, and then builds a bulge or otherwise
increases its central mass, it would tend to move upwards
in the plot, into the regime where 2-armed morphologies are favored.
Whether a spiral galaxy has two or more arms may depend in part upon the 
mass in the central core, 
whether this mass is in stars or
dark matter.
For a given disk mass,
galaxies with high density cores are more likely to be
2-armed, 
while galaxies with lower density cores are
more likely to be flocculent or multi-armed.
How the dark matter is distributed in spiral galaxies
is still uncertain.
It is sometimes assumed that baryons dominate the mass
in the inner parts of spiral galaxies 
(e.g., \citealp{1985ApJ...295..305V}).
However,
detailed studies
of the rotation curves 
of nearby spirals
indicate 
that the disks of moderate-mass spirals are sub-maximal
\citep{2013A&A...557A.131M,
2016ApJ...827L..19L,
2018MNRAS.480.2292S},
meaning that dark matter dominates the mass
even in the cores.  
In higher mass spirals, however, 
the dark matter contribution in the core appears to become
less important, and the disks are near-maximal
\citep{2013A&A...557A.131M,
2016ApJ...827L..19L,
2018MNRAS.480.2292S}.
Although
the spatial distribution of
dark matter in the inner parts of spiral galaxies 
is quite uncertain \citep{2026MNRAS.548ag531P}, 
the distribution of dark matter 
in the inner regions of low mass spirals
may trace that of the baryons
\citep{2016ApJ...827L..19L}.
If this is the case,
this suggests that low mass spirals with larger stellar bulges 
may have proportionally
larger dark matter concentrations in their cores, favoring 2-armed
morphologies.
This contribution from dark matter to the central mass 
may move the boundary between
2-armed and flocculent galaxies downwards in 
Figure 
\ref{fig:spiral_evolution},
causing low mass 2-armed galaxies to have relatively low 
concentrations in optical images.
An increasing M$_{disk}$/M$_{halo}$ ratio with increasing
stellar mass may be responsible for
the bend in the concentration vs.\ log M* relation,
and the threshold at log (M*/M$_{\sun}$) = 10.5 may correspond
to the point at which the disk becomes maximal.
We suggest that two parameters control
the arm count in spiral galaxies: the relative mass of the disk and halo,
and the central mass density.
If the mass distribution is centrally-concentrated, 
it would favor a dominant 2-armed structure.  In contrast, galaxies with
lower central mass densities will be either multi-armed or flocculent,
depending upon the relative
disk vs.\ halo mass.

To test these ideas, it is important to properly map the C-log M*
plot shown in
Figure 
\ref{fig:spiral_evolution}
and better determine the boundaries between the different
arm classes.
This requires more complete samples
of spirals that also include flocculent galaxies.
It also requires better methods for counting the numbers of 
arms in galaxies (see Section 
\ref{sec:disc_2arm_vs_3arm}).
It would also be valuable to use rotation curves
to measure the dynamical mass distributions in 
galaxies of different arm classes and different stellar masses.
Do the rotation curves of 2-armed and 3-armed galaxies differ?
Do 2-armed galaxies have larger shear than 3-armed galaxies?
For the same stellar mass, are 3-armed galaxies more likely
to have maximal disks than 2-armed galaxies? 
The answers to these questions will shed light
on the nature of spiral arms in galaxies.

\subsection{Bulge Growth?} \label{sec:bulge_growth}

When comparing 2-armed galaxies in our two lowest
redshift ranges, 0.2 $<$ z $\le$ 0.4
and 0.4 $<$ z $\le$ 0.6, 
the galaxies in the lower redshift
range have higher observed concentrations for the same mass. 
Most of this difference can be accounted for by 
morphological K-corrections plus a bias due to 
lower effective resolution at larger
distances.  We cannot, however, rule out bulge growth 
as a contributing factor.

According to IllustrisTNG cosmological simulations,
bulges have grown significantly between z $\sim$ 2
and z = 0
\citep{2019MNRAS.487.5416T}.
The observational evidence for recent bulge growth
in spiral galaxies, however, is mixed. 
An analysis of 
stellar populations
in nearby spirals using 
Calar Alto Legacy Integral Field Area (CALIFA)
data suggests that bulges formed
early and did not evolve much over time
\citep{2021MNRAS.504.3058M}.
This is consistent with models in which 
bulges form in the early Universe, via
disk instabilities, clump migration, and other processes 
in gas-rich disks
\citep{2011ApJ...741L..33B, 2009ApJ...703..785D, 2014MNRAS.438.1870D,
2015MNRAS.450.2327Z}.
On the other hand, 
bulge-disk decompositions led 
\citet{2017ApJ...840...79S}
to conclude that bulge masses have increased by a factor of two since
z $\sim$ 1, while
\citet{2014A&A...564L..12T}
concluded that the luminosities of bulges have grown by 30\% since
z $\sim$ 0.8.

Several properties of galaxies
that might affect the C-log M* 
relation are known to vary with redshift.  
These include
the fraction of galaxies 
with high S\'ersic index radial light profiles
\citep{2012ApJ...753..167B},
the 
SFR-vs-M* relation
\citep{2014ApJS..214...15S, 
2023MNRAS.519.1526P},
the size-M* relation 
\citep{2014ApJ...788...28V},
and the 
stellar-mass-to-halo-mass function
\citep{2020A&A...634A.135G}.
Future studies of the C-log M* relations for different arm classes 
are needed
using the full Euclid sample, 
better spiral classification 
methods, improved photometric redshifts,
Euclid-specific redshift bias corrections,
and detailed bulge/disk decompositions, 
to better address the question of 
bulge growth over time.

\subsection{2-Armed vs.\ Grand Design, 
3-Armed vs. Multi-Armed: Comparison to Previous Studies} 
\label{sec:disc_2arm_vs_3arm}

The ratio of the number of 2-armed galaxies to the number of 3-armed
galaxies in 
our Euclid Zoobot study
and in the 
\citet{2016MNRAS.461.3663H}
Galaxy Zoo 2 study
are similar to each other, but starkly 
different from earlier
by-eye classifications by experts
of the ratio of grand design to multi-armed galaxies.
Some statistics from earlier studies are summarized in 
Table \ref{tab:GD_MA_F},
along with details about the sample selection for each
study.
\citet{1987ApJ...314....3E}
used 
photographic images from the Palomar Observatory Sky Survey (POSS)
and high resolution atlas photographs to determine arm classes for 
galaxies
selected from the 
Second Reference Catalogue
of Bright Galaxies (RC2; 
\citealp{1976RC2...C......0D}).
The 
\citet{2013JKAS...46..141A}
study used SDSS images for 
a volume-limited, absolute-magnitude-limited sample of galaxies.
\citet{2015ApJS..217...32B}
classified a distance-limited, blue magnitude-limited 
set of 1114 spiral galaxies into arm classes using Spitzer 3.6 $\mu$m images.
\citet{2024AJ....168..264W}
determined arm classes for 
5093 galaxies
selected from a list of blue spirals published by 
\citet{2010MNRAS.405..783M}.
The statistics in
Table \ref{tab:GD_MA_F}
are provided in two forms: first, by giving percentages
in all three arm classes (grand design,
multi-armed, and flocculent), and second, by omitting the flocculent
galaxies and simply listing the ratio of grand design to multi-armed
galaxies.

Table \ref{tab:GD_MA_F} shows
considerable variations between these published studies.
Part of this variation is due to
sample selection.  The masses
of the 
galaxies is an important factor,
since galaxies with different arm classes tend to have different masses.
For example, the 
\citet{2015ApJS..217...32B}
sample is distance-limited,
so is more biased towards lower mass galaxies than SDSS-selected samples,
and thus contains a larger proportion of flocculent galaxies.
Another factor that probably
contributes to the differences in 
Table \ref{tab:GD_MA_F} is
the specific criteria used by the researchers to classify the galaxies.
For example, 
\citet{2024AJ....168..264W}
categorize only 3\% of their galaxies
as grand design,
the smallest fraction of grand design galaxies
in
Table \ref{tab:GD_MA_F}.
They use a very strict definition of grand design:
``two long and continuous arms with strong symmetry" and
``devoid of branching along the arms".
They state that with the improved resolution and sensitivity of the
SDSS vs.\ earlier studies, they can detect bifurcations and
discontinuities not previously seen, causing galaxies
to be classified as multi-armed rather than grand design.
Furthermore, they 
state that when a galaxy has only two arms in their
inner region which "spread out" in the outer disk,
they classify it as "multi-armed".
They note that 
more than half of the galaxies that they place in their 
"multi-armed" category have only two arms in their interiors.
The 
\citet{2024AJ....168..264W}
selection criteria for grand design galaxies may favor galaxies
that have recently suffered strong tidal interactions, i.e., 
the 
\citet{2015ApJS..217...32B}
`extreme spirals'.
These may be a small subset of all 2-armed galaxies.

In Table \ref{tab:GD_MA_F}
we also list the
percentage of galaxies classified as 2-armed in 
\citet{2016MNRAS.461.3663H}
and in our Euclid study, 
compared to the sum of the percentages identified
as 3-armed, 4-armed, and 5+-armed (collectively assumed to be 
multi-armed galaxies).  We also provide the percentages
of
`can't-tell' spirals
(possibly flocculent galaxies).
Our Euclid percentages agree well with the 
\citet{2016MNRAS.461.3663H}
GZ2 SDSS fractions.
However, 
the fraction of galaxies placed in the flocculent arm class by
experts are two to three times higher than the 
proportion of galaxies classified by the GZ2 
volunteers/Euclid Zoobot as can't-tell.
The most likely explanation for this is that 
many flocculent galaxies are omited from the spiral category
by 
the GZ viewers/Zoobot. 

Independent of the flocculent/cant-tell galaxies,
the ratio of 2-armed/3+4+5+-armed galaxies in the SDSS GZ2
study and the Euclid Zoobot database is at least four times
larger than the ratio of grand design/multi-armed galaxies 
in the earlier studies.
Clearly, a 
classification as 2-armed by GZ2/Zoobot is not identical
to a classification as grand design by an expert.
One reason for this difference
is that 
many galaxies 
have two arms in their inner disks, which branch into
multiple arms further out in the disk.
In a careful by-eye classification by experts,
this branching may be noticed, and the galaxy placed
in the multi-armed group.   In contrast, 
a volunteer GZ2 participant may take a cursory look and notice
the two arms in the inner regions, and ignore branching
further out in the galaxy. Perhaps the GZ2/Zoobot classifications
are more heavily weighted towards the inner disks of the galaxies,
while expert classifiers may consider the overall arm structure. 
The statistics in 
Table \ref{tab:GD_MA_F} strongly suggest that 
some of the galaxies selected as 2-armed by our initial selection criteria
would be classified as multi-armed by experts. 

In our Euclid study,
our `highly reliable'
selection criteria more cleanly separates 2-armed
galaxies
from multi-armed galaxies compared to our original initial sample selection.
The fact that these `highly reliable' subsets of galaxies
occupy different
locations in concentration-log M* plots demonstrates conclusively
that they
do not come from the same parent sample, but have 
fundamentally different structural and physical properties.
However, the additional filtering used to 
construct
these highly reliable subsets excludes a large number of galaxies
with more uncertain arm counts, making determining the percentage
of spirals in different arm classes very uncertain.

Few galaxies are truly completely
2-armed or completely 3-armed (or completely four-armed), 
but instead there is a continuum
of morphologies, with branches and spurs
being common.   Our highly reliable 2-armed sample is presumably
more weighted towards galaxies in which a 2-armed structure dominates,
compared to the 3-armed sample.
However, spiral arm structure is
more complicated than the three simple categories
of grand design, multi-armed, and flocculent.  
Likewise, the Galaxy Zoo 2/Zoobot arm counts, although
informative, do not provide the complete story.

A more detailed arm identification system with twelve 
categories was introduced
by 
\citet{1982MNRAS.201.1021E}.
The additional information these 12 categories provide may be helpful
in exploring questions about the origin and evolution of spiral
patterns in galaxies.
However, this 12-group method is 
more complicated and time-consuming to
use than the simpler 3-class system, 
and 
would be difficult to implement in an automatic way for large samples
of galaxies.
Given the gigantic numbers of galaxies being
observed by Euclid and soon the Nancy Grace Roman Telescope,
there is a 
need for a 
simple-to-implement
mathematical description of arm structure to
cleanly separate galaxies into subsets
based on arm properties.  
One possibility is a Fourier-type decomposition of images,
a technique previously used for smaller numbers of galaxies
\citep{2004A&A...423..849G,
2009MNRAS.397.1756D,
2011ApJ...737...32E,
2018ApJ...862...13Y,
2020ApJ...900..150Y}.
Determining
Fourier components as a function of the radius of a galaxy
would be helpful in identifying
spiral arm branches and spurs, and could 
pick out galaxies
with two arms in the interior and multiple arms at larger radii.
An alternative classification method that may prove valuable
in studying arm structure is fractal dimensions 
\citep{2025ApJ...991...63S}.

\begin{deluxetable*}{crrrrccc}
\tablecaption{Percentages
of Grand Design, Multi-Armed, and Flocculent Galaxies\label{tab:GD_MA_F}}
\tablewidth{0pt}
\setlength{\tabcolsep}{1pt}
\tablehead{
\colhead{Reference} 
& \colhead{Grand } 
& \colhead{Multi-}
& \colhead{Floc-}
& \colhead{Ratio}
& \colhead{Images}
& \colhead{Source/}
& \colhead{Selection}
\\
\colhead{} 
& \colhead{Design} 
& \colhead{Armed}
& \colhead{culent}
& \colhead{GD/MA}
& \colhead{Used}
& \colhead{Catalog}
& \colhead{Criteria}
\\
}
\startdata
Elmegreen \& Elmegreen 1987 & 13\% & 42\% & 45\% & 0.31 & POSS+Atlas & RC2 & \\
Ann \& Lee 2013 & 19\% & 38\% & 43\% &  0.5 & SDSS & & z $<$ 0.2; M$_{r}$$<$-16.1 \\
Buta et al.\ 2015 & 18\% & 32\% & 50\% & 0.56 & Spitzer  & S$^4$G &  D$<$40 Mpc,m$_{\rm B}$$<$15.5 \\
Wei et al.\ 2024 & 3\% & 59\% & 38\% &  0.05 & SDSS & & blue spirals
\\
\hline \\
Reference & 
2-Armed
& 3+4+5+-
& Can't-
& Ratio 
& Images
& Source/
& Selection\\
& 
& Armed
& \multicolumn{1}{c}{tell}
& 2-Armed/
&
& Catalog
& Criteria\\
& 
& 
& 
& 3+4+5+-
& 
& 
& \\
& 
& 
& 
& Armed
& 
& \\
\hline \\
Hart et al.\ 2016$^1$ & 55\% & 27\%  & 17\%$^2$ & 2.0 & SDSS & GZ2 &  
log M* $\ge$ 10.6 \\
This work$^1$  & 54\% & 24\% & 22\% & 2.2 & Euclid & Zoobot & log M* $\ge$ 10.6 \\
\enddata
\tablenotetext{1}{Excluding galaxies selected as 1-armed.}
\tablenotetext{2}{Estimated from Figure 8 in Hart et al.\ (2016).}
\end{deluxetable*}

Employing
a more detailed morphological classication scheme for
spiral patterns in galaxies may lead to a better
understanding of the various processes involved in creating
and maintaining spiral patterns in galaxies.
Perhaps there are distinctly different 
subclasses of 2-armed spirals, for example,
those with transient m=2 patterns caused by 
disk
instabilities,
vs.\ long-lived 2-armed patterns with
fixed pattern speeds, vs.\ tidally-triggered grand
design spirals, vs.\ spirals associated with bars.   
A deeper look at
subtle 
variations in the spiral structures and the physical
properties of 
galaxies selected by different methods
may lead to useful insights.

\section{Summary} \label{sec:summary}

The 
Euclid telescope 
Quick Data Release 1 
\citep{2025A&A...697A...1E}
has provided the
Astronomical community with several
catalogs of galaxy properties,
containing photometric redshifts,
stellar masses, concentrations, and SFRs
\citep{2025arXiv250315306E}.
For 380,111 galaxies, the Zoobot
foundation model made estimates of the 
expected 
Galaxy Zoo 2
vote fraction
for several morphological traits, including
the number of spiral arms
\citep{2025arXiv250315310E}.

Filtering out galaxies with less reliable parameters, we 
placed each galaxy into an initial arm
count class based on which class has the highest estimated vote
fraction.  We find that 
65\% of the log (M*/M$_{\sun}$) $\ge$ 10.6 spirals 
with 0.175 $\le$ z $<$ 0.275
are 2-armed
and 15\% are 3-armed, if the `can't count number of arms' spirals
are excluded.  These statistics are consistent with the 
\citet{2016MNRAS.461.3663H}
bias-corrected fractions
for SDSS galaxies in the local Universe.
Determining the fraction of spirals of different
arm counts at higher redshifts, z $>$ 0.2, is quite
uncertain, due to sample incompleteness and redshift-dependent
classification biases.  Interestingly, however, 
\citet{2025A&A...700A..42E}
classify 60\% of JWST 
z $\sim$ 1
log (M*/M$_{\sun}$) $\ge$ 10.0 
spirals as 2-armed, similar 
to the 
Euclid z $\sim$ 0.2 Zoobot percentage.
Although differences in classification metrics between
the JWST study and the Zoobot may be 
substantial, as a first tentative attempt to investigate
how spiral fractions vary in the 0.2 $<$ z $\le$ 1 window,
based on these JWST observations
we make the simplifying assumption
that the fraction of 2-armed spirals remains constant in
this redshift range.
Given this assumption, and making an approximate correction
for 
redshift bias, we infer that
the number of 3-armed galaxies 
increases slightly with redshift to $\sim$30\% at z = 1.
This result is quite preliminary, and needs 
confirmation.

Filtering the samples further to only include
`highly reliable' 2-armed galaxies
(vote fraction for 2-arms $\ge$ 0.8), `highly reliable' 3-armed galaxies
(vote fraction for 3-armed $\ge$ 0.5), and `highly reliable' 1-armed
galaxies (vote fraction of 1-armed $\ge$ 0.5), 
we identify 6284 2-armed, 262 3-armed
galaxies, and 188 1-armed galaxies with 0.2 $\le$ z $<$ 1.   
The 1-armed galaxies tend to have lower masses than
the 2-armed galaxies, but for a given stellar mass
1-armed galaxies have similar concentrations and SFRs as
2-armed galaxies.
The
2-armed galaxies have larger concentrations, lower stellar masses,
and lower SFRs, on average, than the 3-armed galaxies.
The lower SFRs for 2-armed galaxies are a consequence of
their lower stellar masses; for the same stellar mass, 2-armed and 3-armed
galaxies have similar SFRs.
These differences between 2-armed and 3-armed
galaxies have been seen before in nearby galaxies 
(e.g., \citealp{2024AJ....168...12S}).
With Euclid we show that the trend with concentration continues out to z = 1,
and the trend for mass and SFR to z = 0.4.

These results are consistent with theoretical models and simulations,
which show that 2-armed galaxies are favored when the bulge is prominent
\citep{1985IAUS..106..513L, 1989ApJ...338...78B,
1989ApJ...338..104B,
2013A&A...553A..77G, 
2015ApJ...808L...8D, 
2016ApJ...826L..21S,
2018MNRAS.481..185M, 2025Galax..13..132P}, and
multi-armed or flocculent morphologies appear in models with
small or no bulge
\citep{2013ApJ...766...34D, 2018MNRAS.477.1451F}.
We surmise that the lack of low mass 3-armed galaxies in our sample
is a consequence of the known trend of 
smaller M*/M$_{halo}$ ratios in lower mass galaxies 
\citep{2010ApJ...717..379B, 2020A&A...634A.135G}.
According to simulations, galaxies with large M*/M$_{halo}$
but small bulges tend to be multi-armed, but galaxies with
small M*/M$_{halo}$ and similarly small bulges will be flocculent 
\citep{2013ApJ...766...34D, 2018MNRAS.477.1451F}.
Since our sample lacks flocculent galaxies, the low C, low M*
regime is only sparsely populated in our study.
Followup studies are needed to better quantify where flocculent
galaxies are located in C-log M* space.

We see a bend in the concentration-log M* relation for 2-armed
galaxies at M* $\approx$ 10$^{10.3}$ M$_{\sun}$. 
High mass 2-armed galaxies show 
significantly larger concentrations than their lower mass counterparts.
In contrast, 3-armed galaxies reveal a linear correlation between
concentration and log M*, without a bend.
We explain 
the bend in the C-log M* relation for 2-armed galaxies as being the
result of several factors: 
1) an increasing M*/M$_{halo}$ ratio with
increasing M*, causing 2-armed morphologies to be favored at
lower stellar concentrations at lower masses,
2) stellar mass
growth being dominated by gas accretion in low mass galaxies,
but by mergers in high mass galaxies, 
and
3) growth of the bulge at high galaxian masses due to mergers or other
processes.

When comparing 2-armed galaxies at z $\sim$ 0.3 and z $\sim$ 0.5, 
for a given stellar mass we see larger concentrations at lower redshifts.
Most of the observed difference in concentration
can be accounted for by a morphological K-correction, plus
lower effective resolution
at higher redshifts.  However, 
we cannot rule out that some of this apparent
increase in concentration is caused by bulge growth over time.
This result needs to be investigated further with larger sample
sizes, better photometric redshifts, and detailed bulge/disk
decompositions of the images.

We compare our Euclid
arm count statistics with earlier statistics
of the relative numbers of grand design, multi-armed, and flocculent
galaxies determined by visual inspection of images by experts.   
We point out systematic selection effects in the samples, and discuss the need
for better methods of quantifying spiral arm structure in galaxies.

{\bf Acknowledgements}
We thank Elena D'Onghia, Bruce Elmegreen, and the anonymous referee for helpful
suggestions that greatly improved this paper.
This research was supported by a grant from the NASA Tennessee Space Grant Consortium.
This work has made use of the Euclid Q1 data from the Euclid mission of the European Space Agency (ESA)
\citep{euclid_dataset}.
We thank the Euclid Consortium for making the Euclid Q1 catalogs available.
The Euclid Consortium acknowledges the European Space Agency and a number of agencies and institutes that have supported the development of Euclid, in particular the Agenzia Spaziale Italiana, the Austrian Forschungsförderungsgesellschaft funded through BMK, the Belgian Science Policy, the Canadian Euclid Consortium, the Deutsches Zentrum für Luft- und Raumfahrt, the DTU Space and the Niels Bohr Institute in Denmark, the French Centre National d’Etudes Spatiales, the Fundação para a Ciência e a Tecnologia, the Hungarian Academy of Sciences, the Ministerio de Ciencia, Innovación y Universidades, the National Aeronautics and Space Administration, the National Astronomical Observatory of Japan, the Netherlandse Onderzoekschool Voor Astronomie, the Norwegian Space Agency, the Research Council of Finland, the Romanian Space Agency, the State Secretariat for Education, Research, and Innovation (SERI) at the Swiss Space Office (SSO), and the United Kingdom Space Agency. A complete and detailed list is available on the Euclid web site www.euclid-ec.org/





%
\facilities{Euclid(VIS)}

\software{astropy \citep{2013A&A...558A..33A,2018AJ....156..123A,2022ApJ...935..167A}
          }





\bibliography{euclid_paper.bib}{}
\bibliographystyle{aasjournalv7}



\end{document}